\documentclass[12pt,a4paper]{article}
\pdfoutput=1
\usepackage[a4paper]{geometry}
\usepackage{graphicx}
\usepackage{amsmath}
\usepackage{amssymb}
\usepackage{float}
\usepackage{appendix}
\usepackage{listings}
\newcommand{\asb}{\bar{\alpha}_s}

\newcommand{\stringa}{\ttfamily\lstinline}
\def\cod#1{{\stringa!#1!}}

\title{\bf Stability of Azimuthal-angle Observables under Higher Order Corrections
in Inclusive Three-jet  Production}
\author{F. Caporale$^1$, F.~G. Celiberto$^{1,2}$, G. Chachamis$^1$, \\ 
        D. Gordo G{\' o}mez$^1$\footnote{`La Caixa'-Severo Ochoa Scholar.}\,\,,  A. Sabio Vera$^{1}$\\ \\
{\small $^1$ Instituto de F{\' \i}sica Te{\' o}rica UAM/CSIC, Nicol{\'a}s Cabrera 15}\\ 
{\small \& Universidad Aut{\' o}noma de Madrid, E-28049 Madrid, Spain.}\\
{\small $^2$ Dipartimento di Fisica, Universit{\`a} della Calabria \&}\\
{\small Istituto Nazionale di Fisica Nucleare, Gruppo Collegato di Cosenza,}\\
{\small I-87036 Arcavacata di Rende, Cosenza, Italy.}
}

\begin{document}

\maketitle 

\abstract
Recently, a new family of observables consisting of azimuthal-angle generalised ratios
was proposed in a kinematical setup that resembles the usual Mueller-Navelet jets but with
an additional tagged jet in the central region of rapidity. Non-tagged minijet activity between the three jets can affect significantly the azimuthal angle orientation of the  jets and is accounted for by the introduction of two BFKL gluon Green functions. Here, we calculate the, presumably, most relevant higher order corrections to the observables by now convoluting the three leading-order jet vertices with two gluon Green functions at next-to-leading logarithmic approximation. The corrections appear to be mostly moderate giving us confidence that the recently proposed observables are actually an excellent way to probe the BFKL dynamics at the LHC. Furthermore, we allow for the jets to take values in different rapidity bins 
in various configurations such that a comparison between our predictions and the experimental data is a straightforward task.
\section{Introduction}

One of the most active fields of research 
 in Quantum Chromodynamics (QCD) 
is the resummation of large logarithms in the center-of-mass energy squared $s$ for processes dominated by
the so-called multi-Regge kinematics (MRK).
To account for these logarithms, one can make use of the Balitsky-Fadin-Kuraev-Lipatov (BFKL) framework in the leading logarithmic (LLA)~\cite{Lipatov:1985uk,Balitsky:1978ic,Kuraev:1977fs,Kuraev:1976ge,Lipatov:1976zz,Fadin:1975cb} and next-to-leading logarithmic (NLLA) approximation~\cite{Fadin:1998py,Ciafaloni:1998gs}. In inclusive multi-jet production, when the outermost in rapidity jets have
a large rapidity difference, we may assume that the process follows the MRK and therefore, the
BFKL resummation becomes relevant.

\begin{figure}[H]
 \centering
 \includegraphics[scale=0.5]{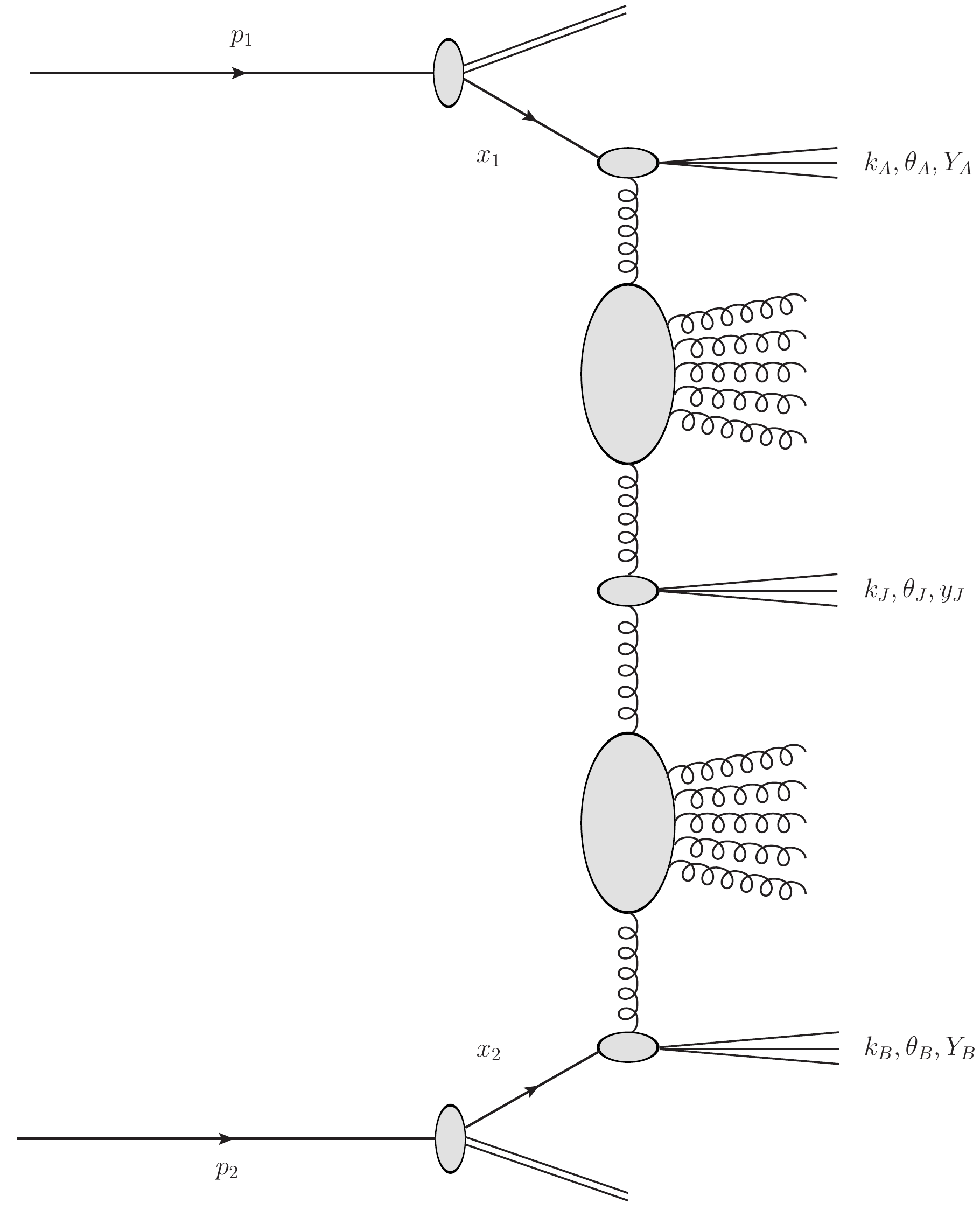}
 \caption[]
 {Inclusive three-jet production process in multi-Regge kinematics.}
 \label{fig:3j}
 \end{figure}

A classic example is Mueller-Navelet jets~\cite{Mueller:1986ey}, that is, the configuration
in hadronic colliders with two final state jets\footnote{
 Another interesting idea,  suggested in~\cite{Ivanov:2012iv} 
and investigated in~\cite{Celiberto:2016hae}, is
the study of the production of two charged light hadrons,
$\pi^{\pm}$, $K^{\pm}$, $p$, $\bar p$, with large transverse momenta and well
separated in rapidity.} which
are produced with large and similar transverse momenta, $k_{A,B}$,  
and a  large rapidity separation $Y=\ln ( x_1 x_2 s/(k_A k_B))$.
$x_{1,2}$ are the longitudinal momentum fractions of the partons that are adjacent to the jets.
Various works~\cite{DelDuca:1993mn,Stirling:1994he,Orr:1997im,Kwiecinski:2001nh,Andersen:2006pg,Angioni:2011wj,Caporale:2013uva, Caporale:2013sc,Marquet:2007xx,Colferai:2010wu,Ducloue:2013wmi,Ducloue:2014koa,
Mueller:2015ael,Chachamis:2015crx,N.Cartiglia:2015gve}   
addressing the azimuthal\footnote{In this work, we denote the azimuthal angles by 
$\theta$ contrary to the usual
practice that prefers $\phi$.} angle ($\theta$) profile of the two tagged jets, 
suggest the presence of important minijet activity populating the rapidity interval which can
be taken into account by considering a BFKL gluon Green function connecting the two jets.
However, it was shown that the azimuthal angle behaviour of the tagged
jets is strongly contaminated by collinear effects~\cite{Vera:2006un,Vera:2007kn}, 
that have their origin at the  $n=0$ Fourier component in $\theta$ of the BFKL kernel. This dependence is mostly canceled if ratios of projections on azimuthal angle observables~${\cal R}^m_n = \langle \cos{(m \, \theta)} \rangle / \langle \cos{(n \, \theta)} \rangle$~\cite{Vera:2006un,Vera:2007kn} (where $m,n$ are integers and $\theta$ is the azimuthal angle between the two tagged jets) are studied.  It also seems that these offer a clearer signal of BFKL effects than the usual predictions for the behaviour of the hadron structure functions $F_{2,L}$ (well fitted within  NLL approaches~\cite{Hentschinski:2012kr,Hentschinski:2013id}). The confrontation of different NLLA
calculations for these ratios ${\cal R}^m_n$~\cite{Ducloue:2013bva,Caporale:2014gpa,Caporale:2014blm,Celiberto:2015dgl,Celiberto:2016ygs} against experimental data at the LHC has  been successful.

Recently, we proposed new observables for processes at the LHC that may be considered as a generalisation
of the Mueller-Navelet jets. These processes are inclusive three-jet~\cite{Caporale:2015vya,Caporale:2016soq}
and four-jet production~\cite{Caporale:2015int,Caporale:2016xku} with the outermost jets widely separated in
rapidity $Y$, whereas any other tagged jet is to be found in more central regions of the detector.
The main idea behind
all this effort is that we need more exclusive final states in order to be able to address a number
of theoretical issues, {\it e.g.} what is the optimal way to implement the running of the strong coupling or
could one speak about saturation effects at present energies, etc.

Investigating more exclusive
final states (with more than two jets) although more challenging on a technical level, allows for
more complex observables to be defined so that one can finally choose those that encapsulate the essence
of these features of MRK that are distinct in the BFKL dynamics only.
In the remaining of this paper, we will focus only on inclusive three-jet production.
\begin{figure}[H]
 \hspace{-1.2cm}
 \includegraphics[scale=0.4]{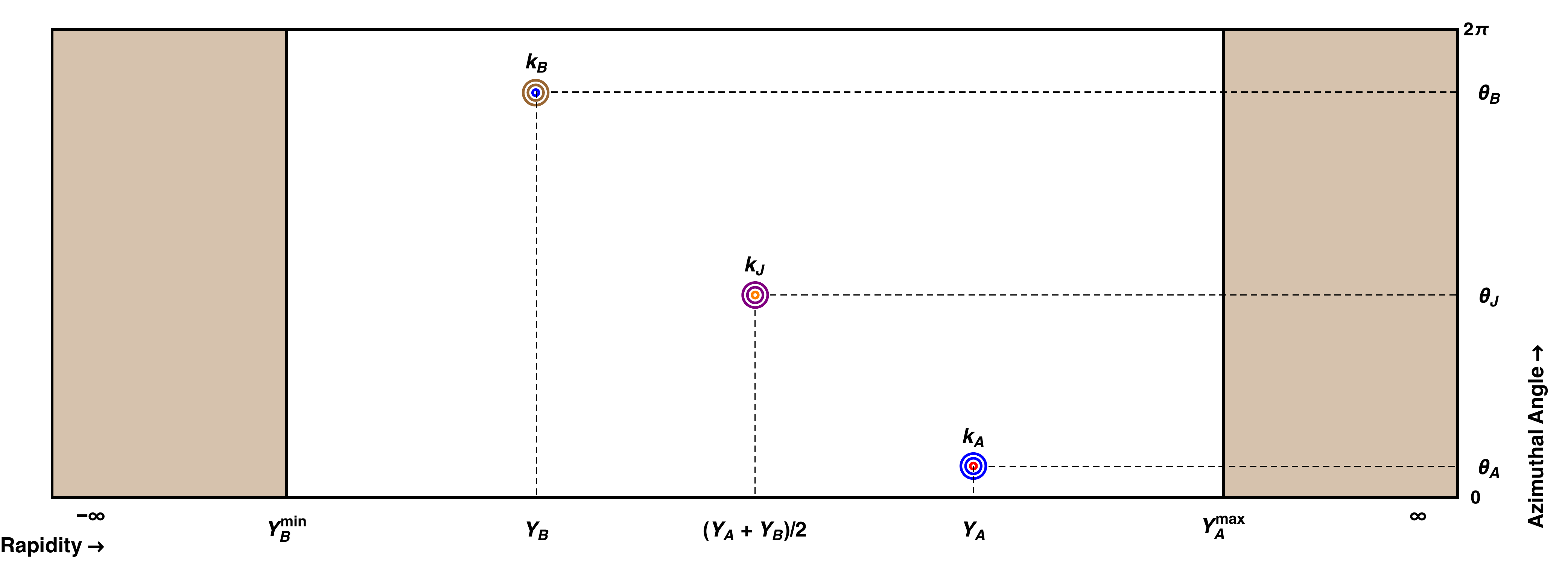}
 \caption[]
 {A primitive lego plot 
 depicting a three-jet event. $k_A$ is a forward jet with 
 large positive
 rapidity $Y_A$ and azimuthal angle $\theta_A$, $k_B$ is a forward jet with 
 large negative
 rapidity $Y_B$ and azimuthal angle $\theta_B$ and
 $k_J$ is a central jet with 
 rapidity $y_J$ and azimuthal angle $\theta_J$. The fade-brown areas to the left and right
 highlight the regions
 in rapidity that are not covered by the standard detectors.
 }
 \label{fig:lego1}
 \end{figure}

The key idea presented in~\cite{Caporale:2015vya} was to get theoretical predictions for the 
partonic-level ratios 
\begin{eqnarray}
{\cal R}^{M N}_{P Q} =\frac{ \langle \cos{(M \, \theta_1)} \cos{(N \, \theta_2)} \rangle}{\langle \cos{(P \, \theta_1)} \cos{(Q \, \theta_2)} \rangle} \, , 
\label{Rmnpq}
\end{eqnarray}
where $\theta_1$ is the azimuthal angle difference between the first and the second (central) jet, while,
$\theta_2$ is the azimuthal angle difference between the second  and the third jet.

In~\cite{Caporale:2016soq}, we presented  a first  phenomenological analysis at LLA
for the respective hadronic-level ratios $R^{M N}_{P Q}$.
These were obtained after using collinear factorization to produce the two most forward/backward jets
and convoluting the 
partonic differential cross section, which follows the BFKL dynamics, with collinear parton distribution functions included in the forward ``jet vertex"~\cite{Caporale:2012:IF,Fadin:2000:gIF,Fadin:2000:qIF,Ciafaloni:1998kx,Ciafaloni:1998hu,Bartels:2001ge,Bartels:2002yj}. In addition, 
the two Mueller-Navelet jet-vertices were linked
with the centrally produced jet via two BFKL gluon Green functions. Finally, we
integrated over the momenta of all produced jets, using actual LHC experimental cuts.

Our predictions in~\cite{Caporale:2016soq}, although may in principle be directly compared to
experimental data once these are available, do not resolve two issues:
(I) They do not offer any estimate of the 
theoretical uncertainty that comes into play once higher order corrections are considered.
(II) Since we restricted the central jet to be produced in the middle of the rapidity
interval between the outermost jets, one could possibly raise concerns of whether 
a experimental analysis following  the kinematical setup used
in Ref.~\cite{Caporale:2016soq} is possible at all. Here, we address both of these issues. 

To that end, regarding issue (I), one needs to calculate higher order corrections for the ratios at partonic-level. This comprises of two steps: considering NLLA corrections to the BFKL kernel and NLO corrections to the jet vertices.  However, although the corrections to the jet vertices may be in general significant, we expect them not to affect much the azimuthal angle characteristics of the jets which are driven mostly by the  minijet activity in the rapidity intervals between the jets. 
Demanding three tagged jets along with central minijets leaves little room for higher order
real emission activity near the jet vertices. We expect that the higher order virtual corrections to the vertices
may be interpreted as K-factor corrections which would cancel out in our observables since we consider ratios.
We have argued previously that the
minijet activity is accounted for by the introduction of the two gluon Green functions.
Large corrections from LLA to NLLA for the gluon Green function, which is actually
a usual outcome in many BFKL-based calculations, could potentially
have a strong impact on the ratios and this at any rate needs to be assessed. Therefore,
in this work we work with NLLA\footnote{Note that from now on,  we will refer to the
results with NLLA gluon Green functions and LO vertices as NLLA results.}
 gluon Green functions and LO jet vertices.

The answer to issue (II) is, naturally,
positive since allowing for the central jet to live in a rapidity range instead of a single point, as long as this
range is located generally in the middle of the rapidity interval between the outermost jets,
does not affect the values of the generalised rations
in Eq.~\ref{Rmnpq}, as was shown in~\cite{Caporale:2015vya}.
Nevertheless, to avoid any confusion and to have a complete study, 
in the present work we are also considering cases in which the central jet
lives in a rapidity bin of unit width, while the central value of the bin may vary.

Other potential sources of uncertainty could be due to the particular PDF sets one uses.
One can still argue though that
the uncertainty due to different PDF sets does not need to be ascertained before
one has gauged how large are the full beyond the LLA corrections to the partonic-level ratios, since
it will be overshadowed by the latter. Indeed, from first tries we see no significant
difference in the results when we work with different PDF sets and therefore
we do not offer any dedicated analysis on that here.

In the bulk of the paper we present  theoretical predictions for the ratios 
$R^{M N}_{P Q}$ at NLLA and we compare these to the LLA ones. 
In particular, in Section 2 we define the computational framework and
our notation for the LLA and NLLA calculations. 
In Section 3 we present results for $R^{12}_{22}$, $R^{12}_{33}$ and $R^{22}_{33}$
as a function of the rapidity distance $Y$ between the outermost jets while the central jet
is fixed at the middle of this distance, for both $\sqrt{s}=7$ and $\sqrt{s}=13$ TeV colliding energies.
In Section 4, we present the LLA and NLLA results for $R^{12}_{22}$, $R^{12}_{33}$ and $R^{22}_{33}$
while the central jet is allowed to take values in the rapidity bin $[-0.5,0.5]$. The results are plotted
again as functions of the rapidity interval $Y$ between the outermost jets for  
$\sqrt{s}=7$ and $\sqrt{s}=13$  TeV.
In Section 5, we do not keep $Y$ fixed at any certain value, instead, we allow for the
forward jet to be in the rapidity interval $[3,4.7]$, for the backward one to be 
in the symmetric rapidity interval $[-4.7,-3]$ while the rapidity of the central jet takes again values 
in a bin of unit 1. The central value of the bin though, may now take five different
values, namely, $\{-1, -0.5, 0, 0.5, 1\}$
and we plot both the LLA and NLLA results for $R^{12}_{22}$, $R^{12}_{33}$ and $R^{22}_{33}$
as a function of that central value, again for $\sqrt{s}=7$ and $\sqrt{s}=13$ TeV.
We finish our work with Conclusions and Outlook.
 
\section{Hadronic inclusive three-jet production in multi-Regge kinematics}

The process under investigation (see Figs.~\ref{fig:3j} and~\ref{fig:lego1})
is the production of two forward/backward jets, both characterized by high transverse momenta $\vec{k}_{A,B}$ and well separated in rapidity, together with a third jet produced in the
central rapidity region and with possible associated minijet production. This corresponds to 
\begin{eqnarray}
\label{process}
{\rm proton }(p_1) + {\rm proton} (p_2) \to 
{\rm j}(k_A, Y_A) + {\rm j}(k_J, y_J) + {\rm j}(k_B,Y_B)  + {\rm minijets}\;,
\end{eqnarray}
where ${\rm j}(k_A, Y_A) $ is the forward jet with transverse momentum $k_A$
and rapidity $Y_A$, ${\rm j}(k_B, Y_B) $ is the backward jet with transverse momentum $k_B$
and rapidity $Y_B$ and ${\rm j}(k_J, y_J) $ is the central jet with transverse momentum $k_J$
and rapidity $y_J$.

In collinear factorization the cross section for the process~(\ref{process}) reads
\begin{align}
\label{dsigma_pdf_convolution}
 & \frac{d\sigma^{3-{\rm jet}}}
      {dk_A \, dY_A \, d\theta_A \, 
       dk_B \, dY_B \, d\theta_B \, 
       dk_J \, dy_J d\theta_J}  = 
 \\ \nonumber 
 \hspace{1cm}& \sum_{r,s=q,{\bar q},g}\int_0^1 dx_1 \int_0^1 dx_2
 \ f_r\left(x_1,\mu_F\right)
 \ f_s\left(x_2,\mu_F\right) \;
 d{\hat\sigma}_{r,s}\left(\hat{s},\mu_F\right) \;,
\end{align}
where the $r, s$ indices specify the parton types 
(quarks $q = u, d, s, c, b$;
antiquarks $\bar q = \bar u, \bar d, \bar s, \bar c, \bar b$; or gluon $g$),
$f_{r,s}\left(x, \mu_F \right)$ are the initial proton PDFs; 
$x_{1,2}$ represent the longitudinal fractions of the partons involved 
in the hard subprocess; $d\hat\sigma_{r,s}\left(\hat{s}, \mu_F \right)$ 
is the partonic cross section for the production of jets and
$\hat{s} \equiv x_1x_2s$ is the squared center-of-mass energy of the
hard subprocess (see Fig.~\ref{fig:3j}). The BFKL dynamics enters in the cross-section 
for the partonic hard subprocess $d{\hat\sigma}_{r,s}$ in the form of two forward gluon Green functions $\varphi$ to be described in a while. 

Using the definition of the jet vertex in the leading order approximation~\cite{Caporale:2012:IF}, 
we can present the cross section for the process as
\begin{align}
 & \frac{d\sigma^{3-{\rm jet}}}
      {dk_A \, dY_A \, d\theta_A \, 
       dk_B \, dY_B \, d\theta_B \, 
       dk_J \, dy_J d\theta_J} = 
 \nonumber \\ \hspace{1cm}&  
 \frac{8 \pi^3 \, C_F \, \asb^3}{N_c^3} \, 
 \frac{x_{J_A} \, x_{J_B}}{k_A \, k_B \, k_J} \,
 \int d^2 \vec{p}_A \int d^2 \vec{p}_B \,
 \delta^{(2)} \left(\vec{p}_A + \vec{k}_J- \vec{p}_B\right) \,
 \nonumber \\  \hspace{1cm}& \times 
 \left(\frac{N_c}{C_F}f_g(x_{J_A},\mu_F)
 +\sum_{r=q,\bar q}f_r(x_{J_A},\mu_F)\right) \,
 \nonumber \\ \hspace{1cm}& \times
 \left(\frac{N_c}{C_F}f_g(x_{J_B},\mu_F)
 +\sum_{s=q,\bar q}f_s(x_{J_B},\mu_F)\right)
 \nonumber \\ \hspace{1cm}& \times
 \varphi \left(\vec{k}_A,\vec{p}_A,Y_A - y_J\right) 
 \varphi \left(\vec{p}_B,\vec{k}_B,y_J - Y_B\right),
\end{align}
where $N_c$ is the number of colors in QCD and $C_F$ is the Casimir
operator, $C_F = (N_c^2-1)/(2N_c)$. 
In order to lie within multi-Regge kinematics, we have considered the ordering in the rapidity of the produced particles $Y_A > y_J > Y_B$, while $k_J^2$ is always 
above the experimental resolution scale.
$x_{J_{A,B}}$ are the longitudinal momentum fractions
of the two external jets, linked to the respective rapidities 
$Y_{{A,B}}$ by the relation 
$x_{{A,B}} = k_{A,B} \, e^{\, \pm \, Y_{{A,B}}} / \sqrt{s}$. 
$\varphi$ are BFKL gluon Green functions normalized to 
$ \varphi \left(\vec{p},\vec{q},0\right) = \delta^{(2)} \left(\vec{p} - \vec{q}\right)$ 
and $\bar{\alpha}_s$ is defined in  terms of the strong coupling as
 $\bar{\alpha}_s =  N_c/\pi \, \alpha_s \left(\mu_R\right)$.

Building up on the work in Refs.~\cite{Caporale:2015vya,Caporale:2016soq}, 
we study observables for which the BFKL approach will be distinct from other formalisms and also 
rather insensitive to possible higher order corrections. We focus on new quantities 
whose associated distributions are different from the ones which
characterize the Mueller-Navelet case, though still related 
to the azimuthal-angle correlations by projecting the differential cross section
on the two relative azimuthal angles between each external jet
and the central one 
$\Delta\theta_{\widehat{AJ}} = \theta_A - \theta_J - \pi$ and 
$\Delta\theta_{\widehat{JB}} = \theta_J - \theta_B - \pi$ (see Fig.~\ref{fig:lego1}). 
Taking into account the factors coming from the jet vertices, 
it is possible to rewrite the
projection of the differential cross section on the azimuthal angle differences
(Eq.~(7) in Ref.~\cite{Caporale:2015vya} )
in the form
\begin{align}
 \label{lo-nlo}
 & \int_0^{2 \pi} d \theta_A \int_0^{2 \pi} d \theta_B \int_0^{2 \pi} 
 d \theta_J \cos{\left(M \Delta\theta_{\widehat{AJ}} \right)} \,
            \cos{\left(N \Delta\theta_{\widehat{JB}} \right)}\\
 & \hspace{0.5cm} 
 \frac{d\sigma^{3-{\rm jet}}}
      {dk_A \, dY_A \, d\theta_A \, 
       dk_B \, dY_B \, d\theta_B \, 
       dk_J \, dy_J d\theta_J}    = 
 \nonumber \\
 &  \hspace{0.07cm} 
 \frac{8 \pi^4 \, C_F \, \asb^3}{N_C^3} \, 
 \frac{x_{J_A} \, x_{J_B}}{k_A \, k_B} 
 \left(\frac{N_C}{C_F}f_g(x_{J_A},\mu_F) \,
 +\sum_{r=q,\bar q}f_r(x_{J_A},\mu_F)\right) \,
 \nonumber \\ 
 & \times \hspace{0.07cm}
 \left(\frac{N_C}{C_F}f_g(x_{J_B},\mu_F)
 +\sum_{s=q,\bar q}f_s(x_{J_B},\mu_F)\right) \, 
 \sum_{L=0}^{N} 
 \left( \begin{array}{c}
 \hspace{-.2cm}N \\
 \hspace{-.2cm}L\end{array} \hspace{-.18cm}\right)
 \left(k_J^2\right)^{\frac{L-1}{2}}
 \nonumber \\ 
 & \times \hspace{0.07cm}
 \int_{0}^\infty d p^2 \, \left(p^2\right)^{\frac{N-L}{2}} \,
 \int_0^{2 \pi}  d \theta    \frac
 {(-1)^{M+N} \cos{ \left(M \theta\right)} \cos{\left((N-L) \theta\right)}}
 {\sqrt{\left(p^2 + k_J^2+ 2 p k_J \cos{\theta}\right)^{N}}}
 \nonumber \\ 
 & \times \hspace{0.07cm} 
 \varphi^{(LLA,NLLA)}_{M} \left(k_A^2,p^2,Y_A-y_J\right)
 \varphi^{(LLA,NLLA)}_{N} \left(p^2+ k_J^2 + 2 p k_J \cos{\theta},k_B^2,y_J-Y_B\right). \nonumber
\end{align}
In this expression the gluon Green function $\varphi$ is
either at LLA ($\varphi^{(LLA)}$) or at NLLA ($\varphi^{(NLLA)}$)
accuracy. In particular, at LLA we have
\begin{align}
 \varphi^{(LLA)}_{n} \left(k^2,q^2,y\right) \; &= \; 
 2 \, \int_0^\infty d \nu   
 \cos{\left(\nu \ln{\frac{k^2}{q^2}}\right)}  
 \frac{e^{\bar{\alpha}_s  \chi_{|n|} \left(\nu\right) y}}
      {\pi \sqrt{k^2 q^2} }, \label{phinLO}
 \end{align}
while the LLA BFKL kernel  $\chi_{n} \left(\nu\right) $ reads
\begin{align} 
 \chi_{n} \left(\nu\right) \; &= \; 2\, \psi (1) - 
 \psi \left( \frac{1+n}{2} + i \nu\right) - 
 \psi \left(\frac{1+n}{2} - i \nu\right)
\end{align}
and $\psi$ is the logarithmic derivative of Euler's gamma function.

At NLLA we have
\begin{equation}
 \varphi^{(NLLA)}_{n} \left(k^2,q^2,y\right)  
 =2 \int_0^\infty d \nu   
 \cos{\left(\nu \ln{\frac{k^2}{q^2}}\right)}  
 \frac{e^{\bar{\alpha}_s \left(  \chi_{|n|} (\nu) + \bar{\alpha}_s \chi_{|n|}^{(1)}(\nu)  \right) Y}}
      {\pi \sqrt{k^2 q^2} } \, , \label{phinNLO}
\end{equation}
where the NLLA contribution $\chi_{|n|}^{(1)}(\nu)$, calculated in~\cite{Kotikov:2000pm} (see
also~\cite{Kotikov:2000pm2}), can be presented in the form
\begin{equation}
\chi_{n}^{(1)}(\nu)=-\frac{\beta_0}{8\, N_c}\left(\chi_n^2(\nu)-\frac{10}{3}
\chi_n(\nu)-i\chi^\prime_n(\nu)\right) + {\bar  \chi_{n}}(\nu)\, ,
\label{ch11}
\end{equation}
with 
\begin{eqnarray}
-4 \bar \chi_{n}(\nu) &=& \frac{\pi^2-4}{3}\chi_n(\nu)
-6\zeta(3)-\chi_n^{\prime\prime}(\nu) +\,2\,\phi_n(\nu)+\,2\,\phi_n(-\nu) \nonumber\\
&& \hspace{-3cm} +
\frac{\pi^2\sinh(\pi\nu)}{2\,\nu\, \cosh^2(\pi\nu)} \left(
\left(3+\left(1+\frac{n_f}{N_c^3}\right)\frac{11+12\nu^2}{16(1+\nu^2)}\right)
\delta_{n0}
-\left(1+\frac{n_f}{N_c^3}\right)\frac{(1+4\nu^2)\delta_{n2}}{32(1+\nu^2)}
\right),\hspace{1cm}
\end{eqnarray}
and 
\begin{eqnarray}
\phi_n(\nu) &=& \sum_{k=0}^\infty\frac{(-1)^{k+1}}{k+(n+1)/2+i\nu}\left[\psi'(k+n+1)
-\psi'(k+1)\right.\nonumber\\
&&\hspace{-2.2cm}\left.+(-1)^{k+1}(\beta'(k+n+1)+\beta'(k+1)) -\frac{(\psi(k+n+1)-\psi(k+1))}{k+(n+1)/2+i\nu}
\right],
\end{eqnarray}
whereas $4 \beta'(z) = \psi' \left((z+1)/2 \right) -\psi' \left(z / 2\right)$.

In order to make an appropriate choice of the renormalization scale $\mu_R$, we used the Brodsky-Lepage-Mackenzie (BLM) prescription \cite{BLM} which is proven a very successful choice for fitting the data
in Mueller-Navelet studies~\cite{Ducloue:2013bva,Caporale:2014gpa}.
It consists of using the MOM scheme and choosing the scale $\mu_R$ such that the $\beta_0$-dependence of a given observable vanishes.
Applying the BLM prescription leads to the modification of the exponent in Eq.~(\ref{phinNLO}) in the following way:
\begin{equation}
\bar{\alpha}_s \left(  \chi_{|n|} (\nu) + \bar{\alpha}_s \chi_{|n|}^{(1)}(\nu)  \right) Y \,\, \to \,\, 
\bar{\alpha}_s \left(  \chi_{|n|} (\nu) \left(  1+ \frac{\alpha_s}{\pi} T  \right) + \bar{\alpha}_s \chi_{|n|}^{(1)}(\nu)  \right) Y \, ,
\end{equation}
where
\begin{eqnarray*}
T&=&T^{\beta}+T^{\,\rm conf}\;,\\
T^{\beta}&=&-\frac{\beta_0}{2}\left( 1+\frac{2}{3}I \right)\;,\\
T^{conf}&=& \frac{C_A}{8}\left[ \frac{17}{2}I +\frac{3}{2}\left(I-1\right)\xi
+\left( 1-\frac{1}{3}I\right)\xi^2-\frac{1}{6}\xi^3 \right]\;.
\end{eqnarray*}
Here $I=-2\int_0^1dx\frac{\ln\left(x\right)}{x^2-x+1}\simeq 2.3439$ and
$\xi$ is a gauge parameter, fixed at zero.

Following this procedure, the renormalization scale $\mu_R$  is fixed at the value
\begin{equation}
 ( \mu_R^{\rm BLM})^2=k_{A}k_{B}\ \exp\left[\frac{1+4I}{3}+\frac{1}{2} \chi_{n}\left(\nu\right)\right] \, .
\end{equation}
In our numerical analysis we consider two cases. In one, we set $\mu_R=\mu_R^{\rm BLM}$  only in the exponential factor of the gluon Green function $\varphi_n$, while we let the argument of the 
$\bar{\alpha}_s^3$ in Eq.~\ref{lo-nlo} to be at the `natural' scale $\sqrt{k_{A}k_{B}}$, that is,
$\bar{\alpha}_s^3(\sqrt{k_{A}k_{B}})$.
In the second case, we fix 
$\mu_R=\mu_R^{\rm BLM}$ everywhere in Eq.~\ref{lo-nlo}. These two cases lead 
in general to two different but similar values for our NLLA predictions and wherever we present plots
we fill the space in between so that we end up having a band instead of a single curve for the
NLLA observables.
The band represents the uncertainty that comes into play after using the BLM prescription since
there is no unambiguous way to apply it.

The experimental observables we initially proposed are based on the partonic-level
average values (with $M,N$ being positive integers)
\begin{eqnarray}
\label{Cmn}
 {\cal C}_{MN} \, = \,
 \langle \cos{\left(M \left( \theta_A - \theta_J - \pi\right)\right)}  
 \cos{\left(N \left( \theta_J - \theta_B - \pi\right)\right)}
 \rangle && \\
 &&\hspace{-9cm} = \frac{\int_0^{2 \pi} d \theta_A d \theta_B d \theta_J \cos{\left(M \left( \theta_A - \theta_J - \pi\right)\right)}  \cos{\left(N \left( \theta_J - \theta_B - \pi\right)\right)}
 d\sigma^{3-{\rm jet}} }{\int_0^{2 \pi} d \theta_A d \theta_B d \theta_J 
 d\sigma^{3-{\rm jet}} },\nonumber
\end{eqnarray}
whereas,
in order to provide testable predictions for
the current and future experimental data, we introduce 
the hadronic-level values $C_{MN}$ after integrating ${\cal C}_{M,N}$ over the momenta of the tagged jets,
as we will see in the following sections.

From a more theoretical perspective, it is important to have as good as possible perturbative 
stability  in our 
predictions (see~\cite{Caporale:2013uva} for a related discussion). 
This can be achieved by removing the contribution stemming  
from the zero conformal spin, 
which corresponds to the index $n=0$ in Eqs.~(\ref{phinLO}) and~(\ref{phinNLO}).
We, therefore, introduce the ratios
\begin{eqnarray}
\label{RPQMN}
R_{PQ}^{MN} \, = \, \frac{C_{MN}}{C_{PQ}}
\label{RmnqpNew}
\end{eqnarray}
which are free from any $n=0$ dependence, as long as $M, N , P, Q >0$. 
The 
postulate that Eq.~\ref{RmnqpNew} generally describes observables with good perturbative 
stability is under scrutiny in Sections 3, 4 and 5 where we compare LLA and NLLA results.

Before we proceed to our numerical results in the next sections,
 we should give a few details with regard to our numerical computations.
 From all the possible ratios, we have chosen to study the following three:
 $R^{12}_{22}$, $R^{12}_{33}$ and $R^{22}_{33}$. These are enough to
 have an adequate view of how the generic $R^{MN}_{PQ}$ behaves.
We computed $R^{12}_{22}$, $R^{12}_{33}$ and $R^{22}_{33}$ in all cases almost exclusively in
\textsc{Fortran} whereas \textsc{Mathematica} was used mainly for cross-checks.
The NLO MSTW 2008 PDF sets~\cite{MSTW:2009} were used 
and for the strong coupling $\alpha_s$ we chose 
a two-loop running coupling setup 
with $\alpha_s\left(M_Z\right)=0.11707$ and five quark flavours. 
We made extensive use of the integration routine 
\cod{Vegas}~\cite{VegasLepage:1978} 
as implemented in the \cod{Cuba} library~\cite{Cuba:2005,ConcCuba:2015}.
Furthermore, we used the \cod{Quadpack} library~\cite{Quadpack:book:1983}
and a slightly modified version 
of the \cod{Psi}~\cite{RpsiCody:1973} routine.

\begin{figure}[p]
\newgeometry{left=-10cm,right=1cm}
\vspace{-2cm}
\centering

   \hspace{-16.25cm}
   \includegraphics[scale=0.3]{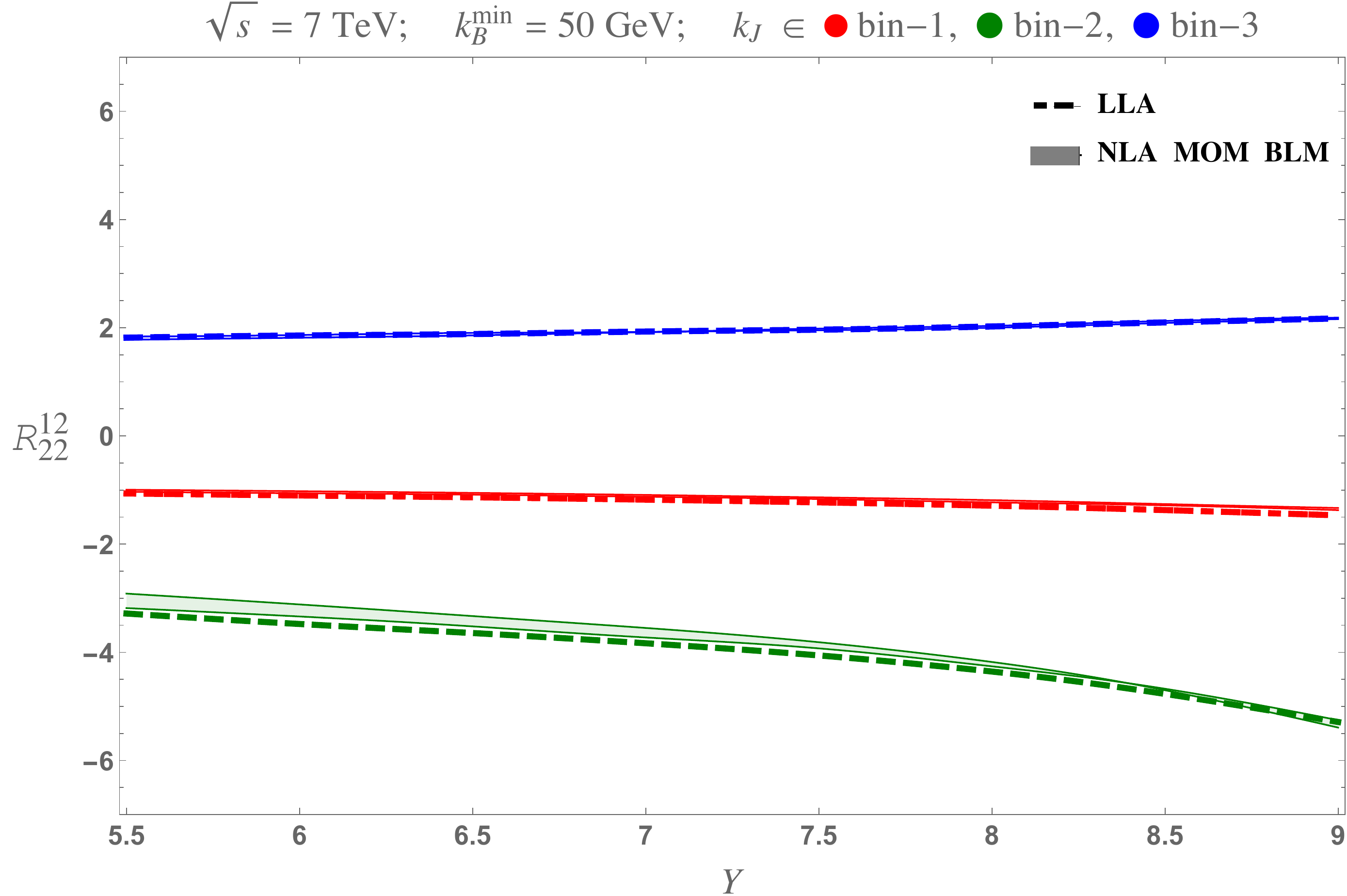}
   \includegraphics[scale=0.3]{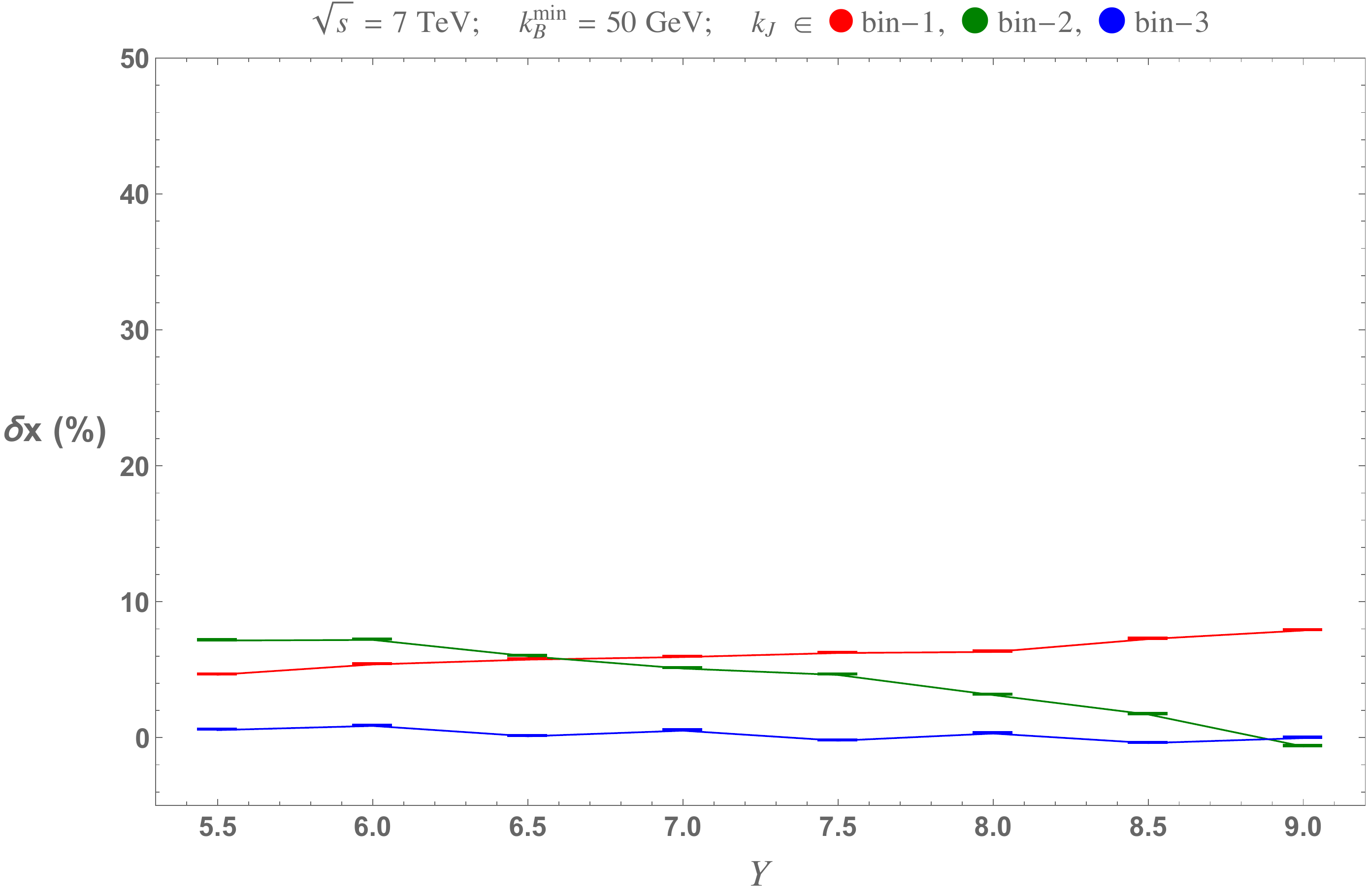}
   \vspace{1cm}

   \hspace{-16.25cm}
   \includegraphics[scale=0.3]{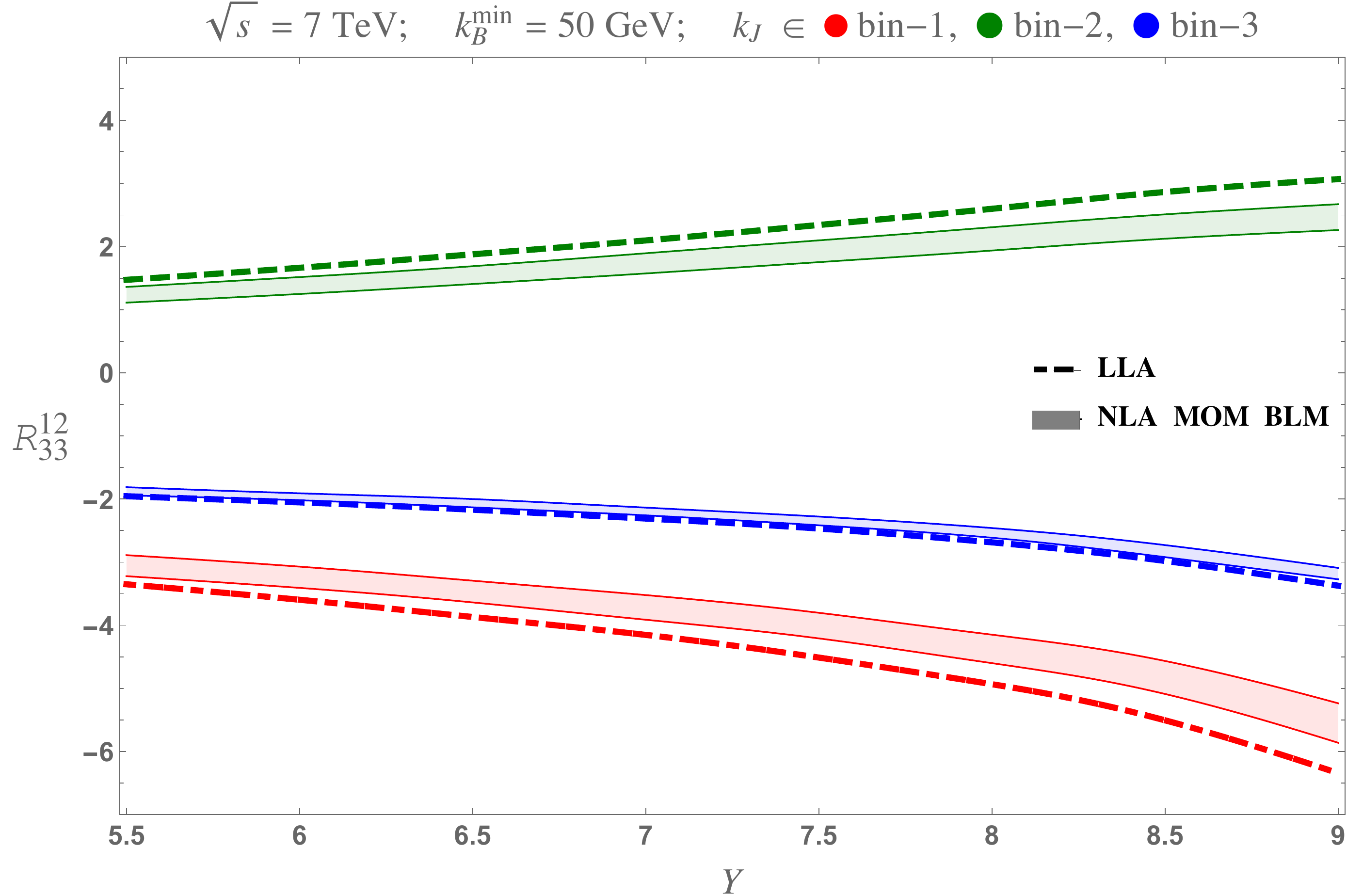}
   \includegraphics[scale=0.3]{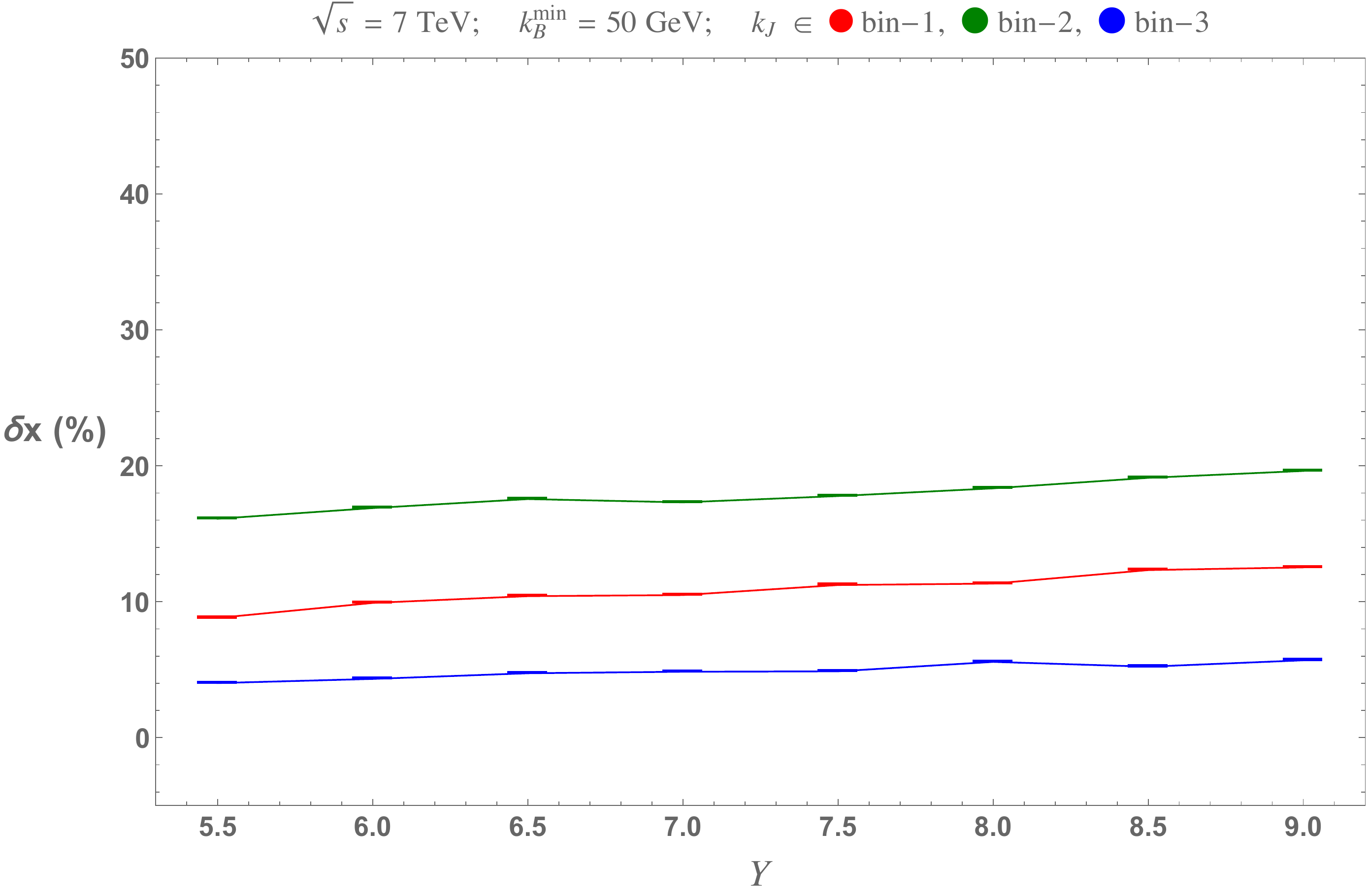}
   \vspace{1cm}

   \hspace{-16.25cm}   
   \includegraphics[scale=0.3]{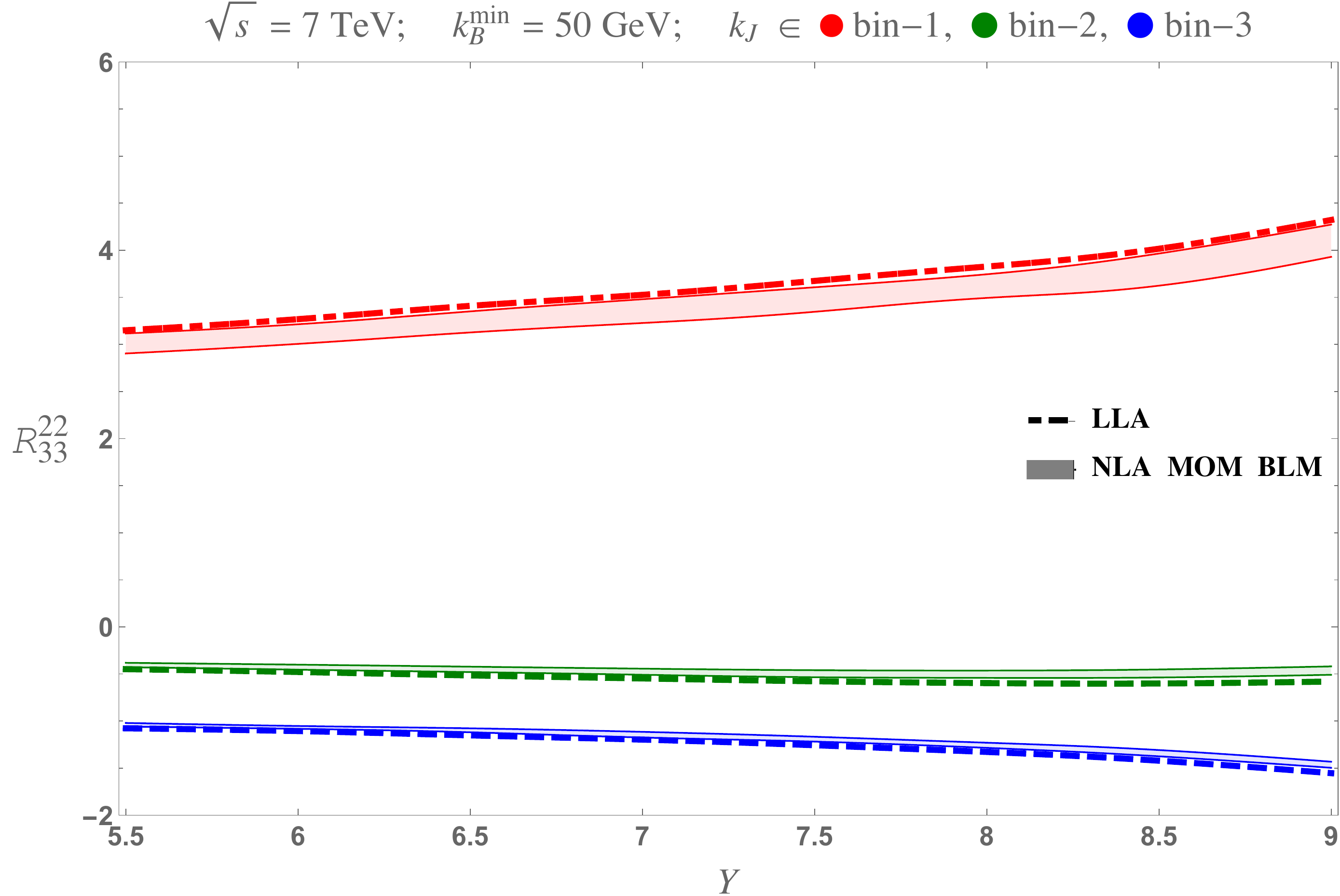}
   \includegraphics[scale=0.3]{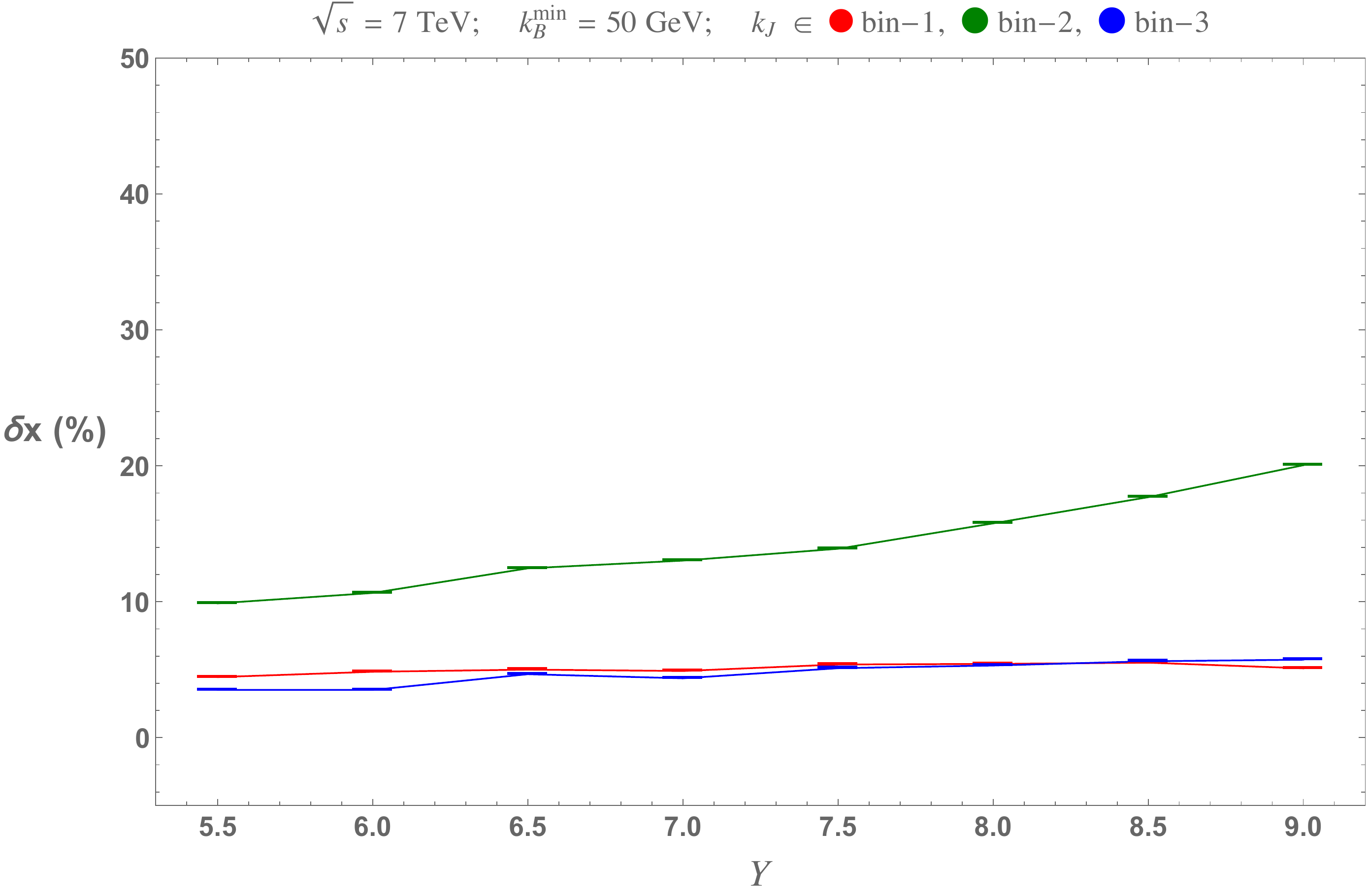}

\restoregeometry
\caption{\small $Y$-dependence of the LLA and NLLA
$R^{12}_{22}$, $R^{12}_{33}$ and $R^{22}_{33}$ at $\sqrt s = 7$ TeV with $y_J$ fixed
 (left) and the relative NLLA to LLA corrections  (right).} 
\label{fig:7-first}
\end{figure}

\begin{figure}[p]
\newgeometry{left=-10cm,right=1cm}
\vspace{-2cm}
\centering

   \hspace{-16.25cm}
   \includegraphics[scale=0.3]{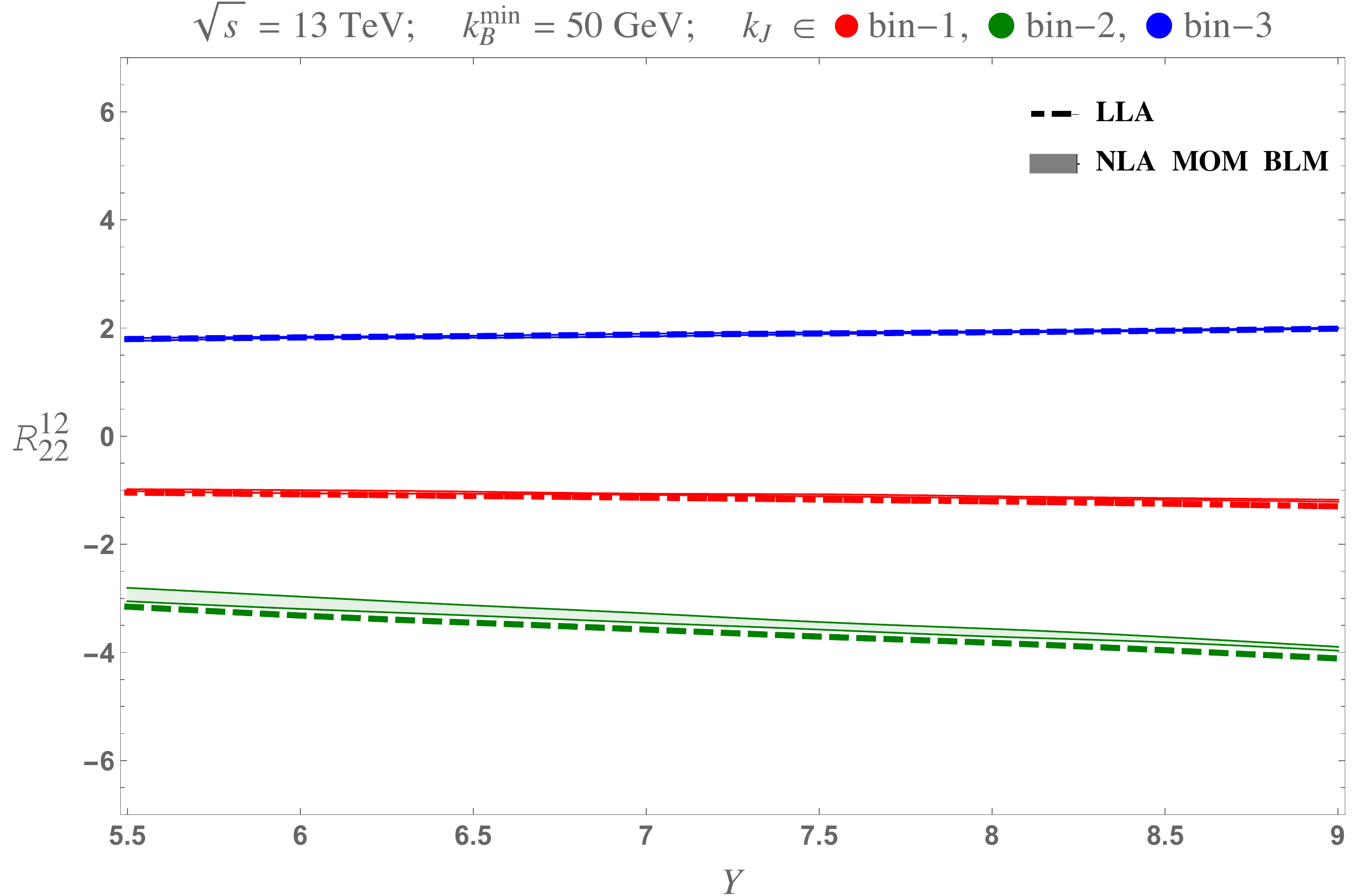}
   \includegraphics[scale=0.3]{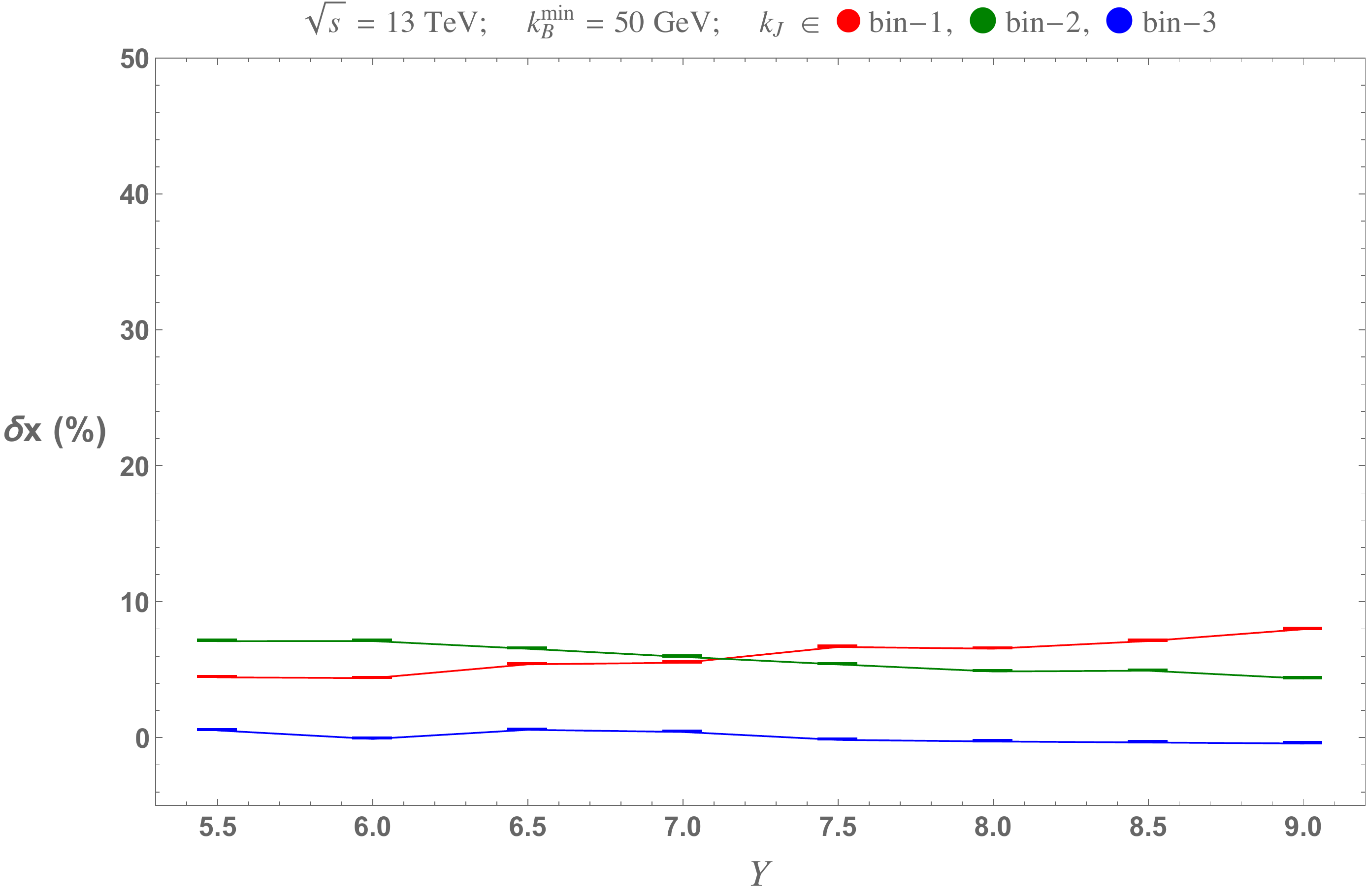}
   \vspace{1cm}

   \hspace{-16.25cm}
   \includegraphics[scale=0.3]{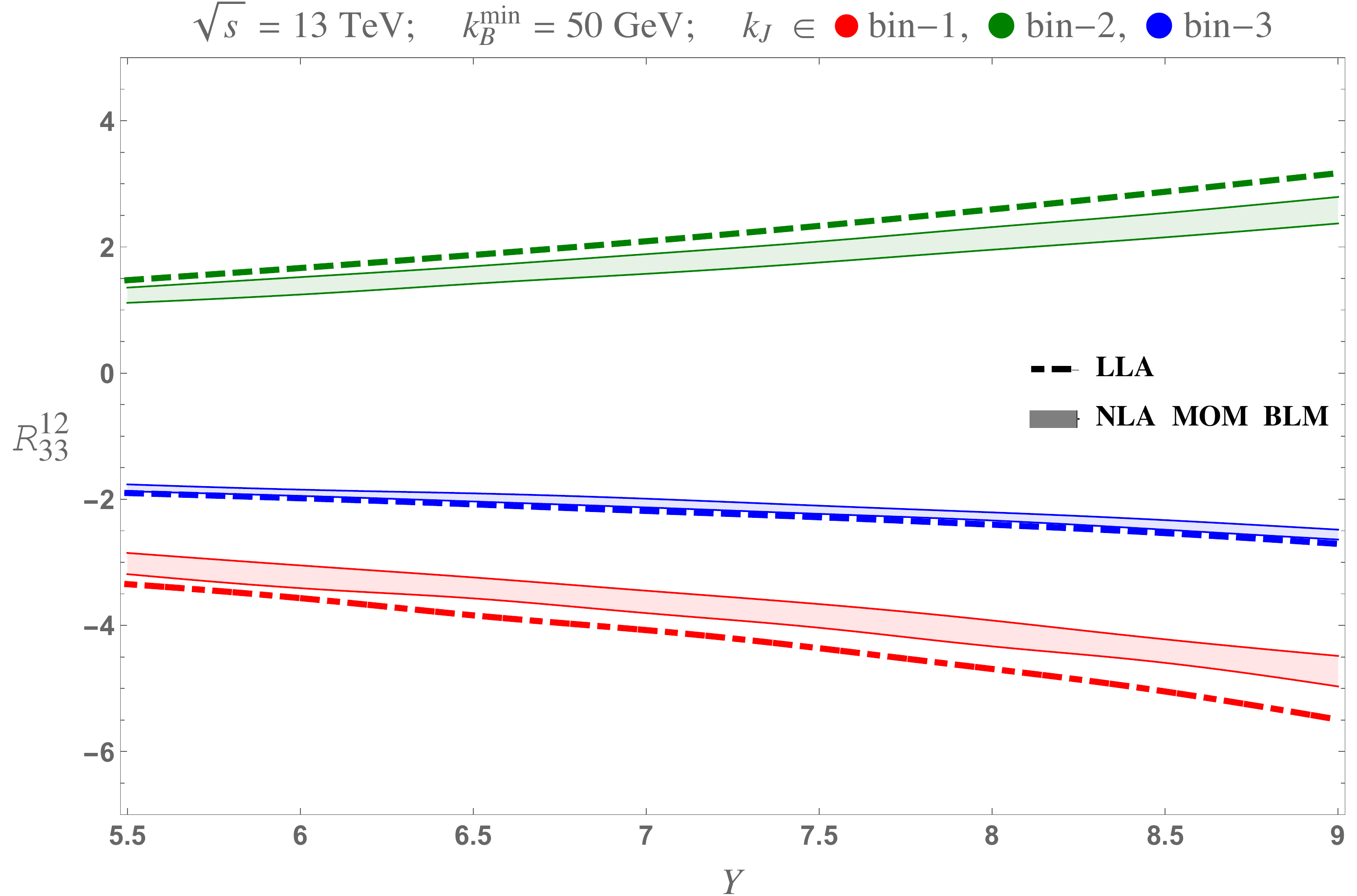}
   \includegraphics[scale=0.3]{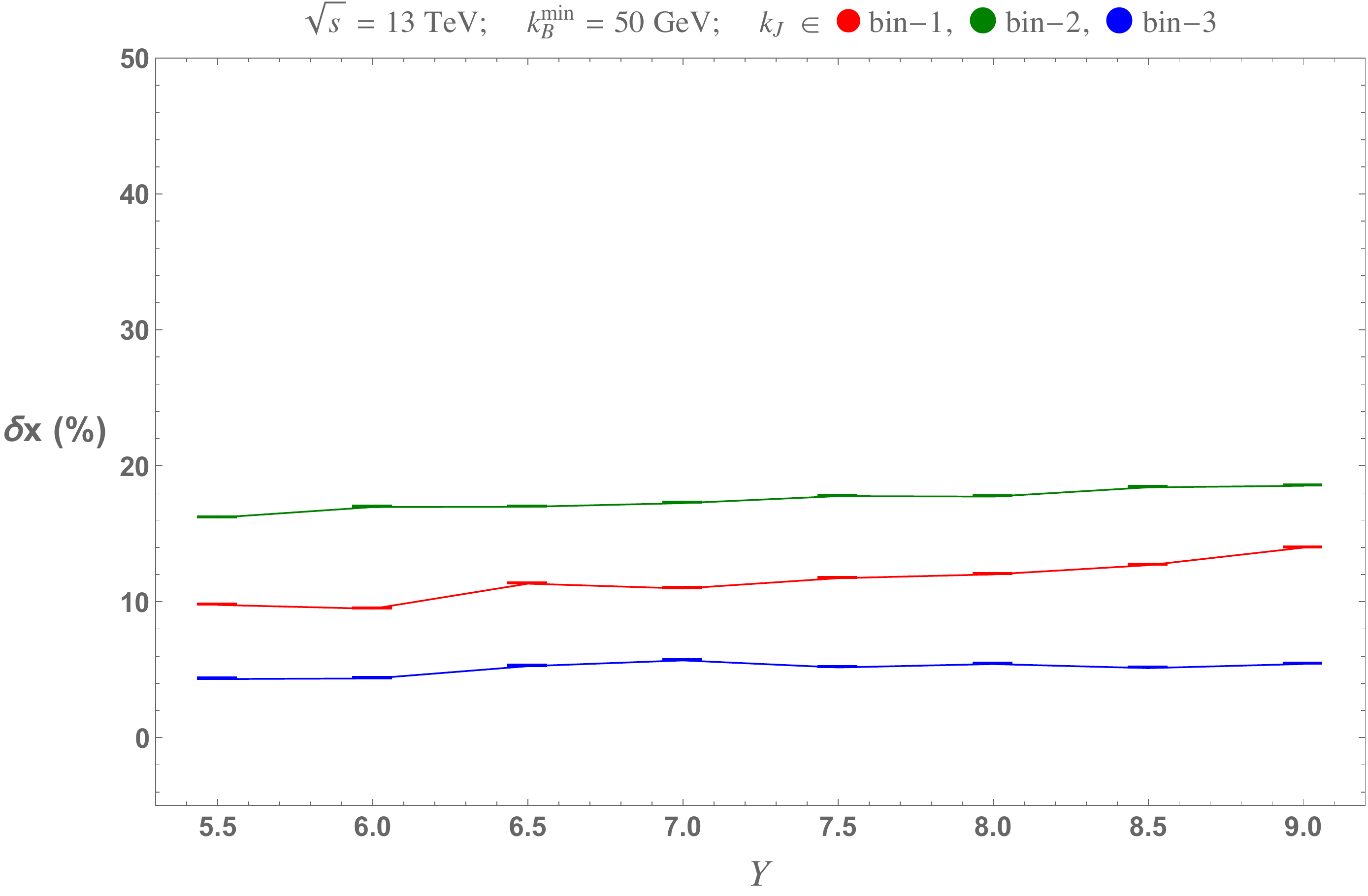}
   \vspace{1cm}

   \hspace{-16.25cm}   
   \includegraphics[scale=0.3]{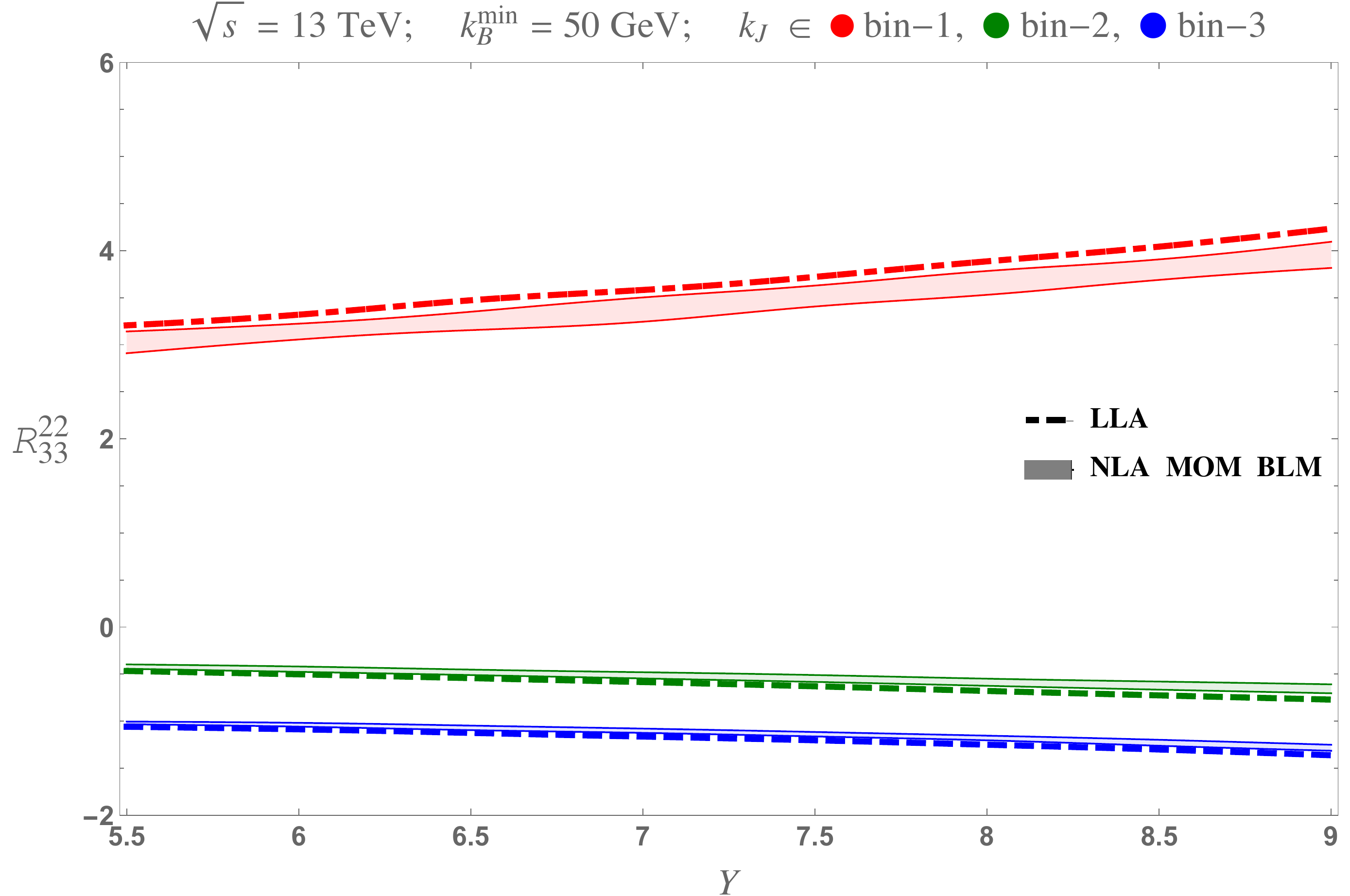}
   \includegraphics[scale=0.3]{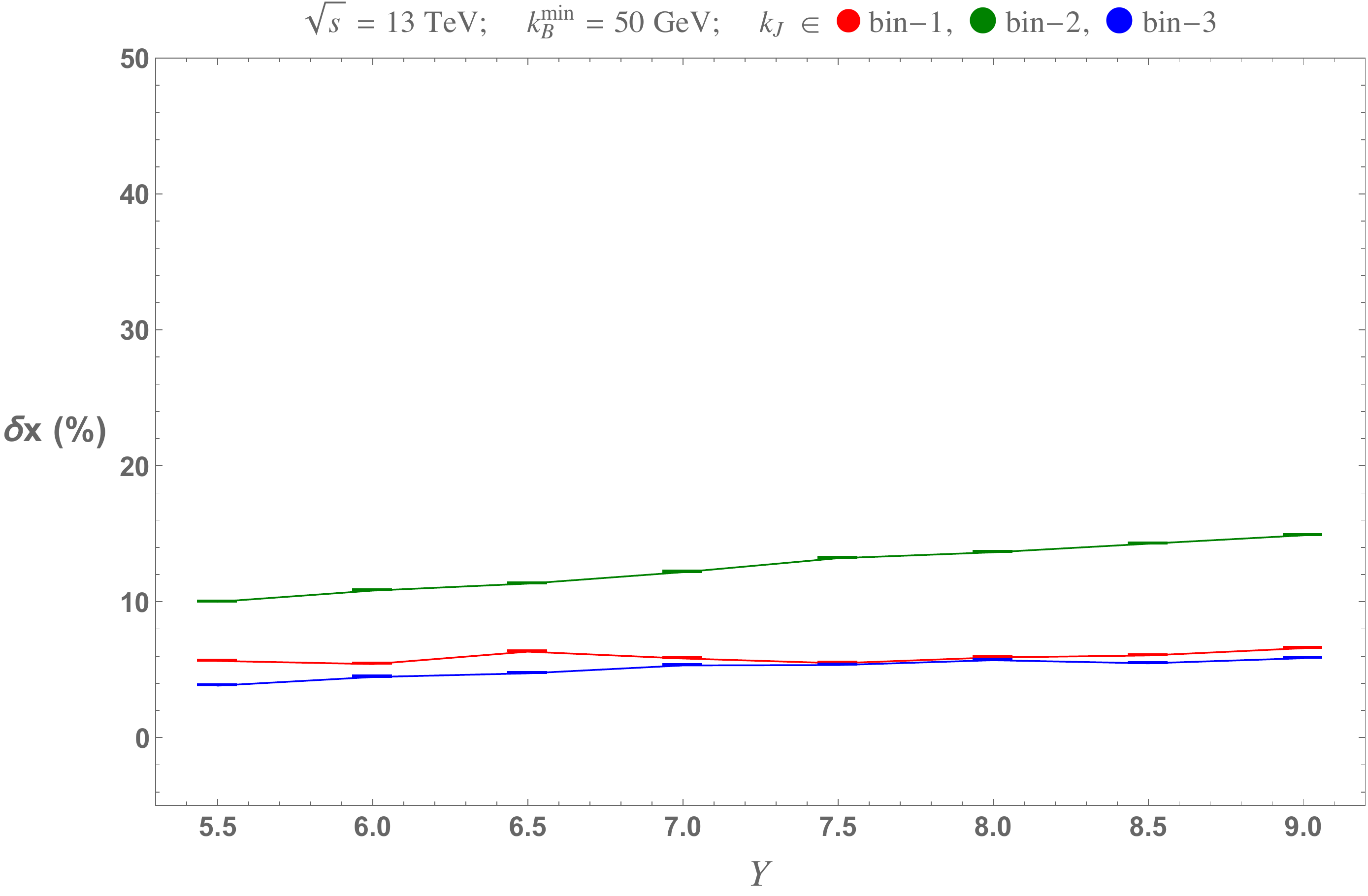}

\restoregeometry
\caption{\small $Y$-dependence of the LLA and NLLA
$R^{12}_{22}$, $R^{12}_{33}$ and $R^{22}_{33}$ at $\sqrt s = 13$ TeV with $y_J$ fixed
 (left) and the relative NLLA to LLA corrections (right).} 
\label{fig:13-first}
\end{figure}

\section{$R^{12}_{22}$, $R^{12}_{33}$ and $R^{22}_{33}$ with the central jet fixed in rapidity}
In this section, we will present results for three generalised ratios, $R^{12}_{22}$, $R^{12}_{33}$ and $R^{22}_{33}$,
assuming that the central jet is fixed in rapidity at $y_J = (Y_A+Y_B)/2$
(see Fig.~\ref{fig:lego1}). 
In particular,
\begin{align}
\label{Cmn_int}
 & 
 C_{MN} =
 \nonumber \\
 &
 \int_{Y_A^{\rm min}}^{Y_A^{\rm max}} \hspace{-0.25cm} dY_A
 \int_{Y_B^{\rm min}}^{Y_B^{\rm max}} \hspace{-0.25cm} dY_B
 \int_{k_A^{\rm min}}^{k_A^{\rm max}} \hspace{-0.25cm} dk_A
 \int_{k_B^{\rm min}}^{k_B^{\rm max}} \hspace{-0.25cm} dk_B
 \int_{k_J^{\rm min}}^{k_J^{\rm max}} \hspace{-0.25cm} dk_J
 \delta\left(Y_A - Y_B - Y\right) {\cal C}_{MN},
\end{align}
where the forward jet rapidity is taken in the
range delimited by $0 < Y_A < 4.7$,
the backward jet rapidity in the range  $-4.7 < Y_B < 0$,
while their difference 
$Y \equiv Y_A - Y_B$ is kept fixed at definite values in the range $5.5 < Y < 9$.

We can now study the ratios  $R_{PQ}^{MN}(Y)$ in Eq.~(\ref{RmnqpNew}) as functions of the 
rapidity difference Y between the most forward and the most backward jets 
for a set of characteristic values of $M, N, P, Q$ and for two different 
center-of-mass energies: $\sqrt s = 7$ and $\sqrt s = 13$ TeV. Since we are integrating over $k_A$ and $k_B$,  
we have the opportunity to impose either symmetric or asymmetric kinematic
cuts, as it has been previously done in  Mueller-Navelet studies.
Here, and for the rest of the paper, we choose to study the asymmetric cut which presents certain
advantages over the symmetric one (see Refs.~\cite{Ducloue:2013wmi,Celiberto:2015dgl}).
To be more precise, we set
$k_A^{\rm min} = 35$ GeV, $k_B^{\rm min} = 50$ GeV,  $k_A^{\rm max} = k_B^{\rm max}  = 60$ GeV
throughout the paper.

In order to be as close as possible to the characteristic rapidity ordering of the multi-Regge kinematics, 
we set the value of the central jet rapidity 
such that it is equidistant to $Y_A$ and $Y_B$ by imposing
the condition $y_J = (Y_A + Y_B)/2$. Moreover, since the tagging of
a central jet permits us to extract more exclusive information from our
observables, we allow three possibilities for the transverse momentum
$k_J$, that is, $20\, \mathrm{GeV} < k_J < 35\, \mathrm{GeV}$ (bin-1),
$35 \,\mathrm{GeV} < k_J < 60\, \mathrm{GeV}$ (bin-2) and
$60\, \mathrm{GeV} < k_J < 120\, \mathrm{GeV}$ (bin-3). Keeping in mind that 
the forward/backward jets have transverse momenta in the  range
$\left[35 \,\mathrm{GeV}, 60 \,\mathrm{GeV}\right]$, restricting the value
of $k_J$ within these three bins allows us to see how the ratio
$R_{PQ}^{MN}(Y)$ changes its behaviour depending on the relative size of the
central jet momentum when compared to the forward/backward ones. Throughout the paper,
we will keep the same setup regarding
bin-1, bin-2 and bin-3 which roughly correspond to the cases
of $k_J$ being `smaller' than, `similar' to and `larger' than
$k_A$, $k_B$, respectively. 

Finally, apart from the functional dependence of the ratios on $Y$
we will also show the relative corrections when we go from LLA to NLLA.
To be more precise, we define 
\begin{eqnarray}
\delta x(\%) = \left(
 \text{res}^{\rm(LLA)} - \frac{\text{res}^{\rm (BLM-1)}+\text{res}^{\rm (BLM-2)}}{2}
 \right) \frac{1}{ \text{res}^{\rm(LLA)}}\,.
 \label{corrections}
\end{eqnarray}
$\text{res}^{\rm(BLM-1)}$ is the BLM NLLA result  for  $\mu_R=\mu_R^{\rm BLM}$ 
only in the gluon Green function while the cubed term of the strong coupling in Eq.~\ref{lo-nlo}
actually reads $\bar{\alpha}_s^3 = \bar{\alpha}_s^3(\sqrt{k_{A}k_{B}})$).
$\text{res}^{\rm (BLM-2)}$ is
the BLM NLLA result  for  $\mu_R=\mu_R^{\rm BLM}$ everywhere in 
Eq.~\ref{lo-nlo}, therefore, $\bar{\alpha}_s^3 = \bar{\alpha}_s^3(\mu_R^{\rm BLM})$,
as was previously discussed in Section 2.

In the following, we present our results for $R^{12}_{22}$, $R^{12}_{33}$ and $R^{22}_{33}$,
with $y_J = (Y_A+Y_B)/2$,
collectively in Fig.~\ref{fig:7-first} ($\sqrt{s} = 7$ TeV) and Fig.~\ref{fig:13-first} ($\sqrt{s} = 13$ TeV),
In the left column  we  are showing plots for  $R^{MN}_{PQ}(Y)$ whereas to the right we are showing 
the corresponding $\delta x(\%)$ between LLA and NLLA corrections.
The LLA results are represented with dashed lines whereas the NLLA ones with a continuous band.
The boundaries of the band are the two different curves we obtain by the two different approaches in
applying the BLM prescription. Since there is no definite way to choose one in favour of the other,
we allow for any possible value in between and hence we end up with a band.
In many cases, as we will see in the following, the two boundaries are so close that the band
almost degenerates into a single curve.
The red curve (band) corresponds to $k_J$  bounded in bin-1,
the green curve (band) to $k_J$  bounded in bin-2 and finally the blue curve (band)
to $k_J$ bounded in bin-3. For  the $\delta x(\%)$ plots we only have three 
curves, one for each of the three different bins of $k_J$.

A first observation from inspecting Figs.~\ref{fig:7-first} and~\ref{fig:13-first}
is that the dependence of the different observables on the rapidity
difference between $k_A$ and $k_B$ is rather smooth.
$R^{12}_{22}$ (top row in Figs.~\ref{fig:7-first} and~\ref{fig:13-first})
at $\sqrt{s} = 7$ TeV and for $k_J$ in bin-1 and bin-3 exhibits an almost linear behaviour with $Y$ both at LLA and NLLA, whereas at $\sqrt{s} = 13$ TeV the linear behaviour is extended also for
$k_J$ in bin-2. The difference  between the NLLA BLM-1 and BLM-2 values
is small, to the point that the blue and the red bands collapse into a single line which in addition
lies very close to the LLA results.
When  $k_J$ is restricted in bin-2 (green curve/band), the uncertainty from applying the BLM
prescription in two different ways seems to be larger.
The relative NLLA corrections at both colliding energies are very modest ranging from
close to $1\%$ for $k_J$ in bin-3 to less than $10\%$ for $k_J$ in the other two bins.

$R^{12}_{33}$ (middle row in Figs.~\ref{fig:7-first} and~\ref{fig:13-first})
compared to $R^{12}_{22}$,
shows a larger difference between BLM-1 and BLM-2 values for $k_J$
in bin-1 and bin-2. The `green' corrections  lower the LLA estimate
whereas the `red' ones  make the corresponding LLA estimate less negative.
The corrections are generally below $20\%$, in particular, `blue' $\sim 5\%$,
`red' $\sim 10\%$ and `green' $\sim 20\%$.

Finally, $R^{22}_{33}$ (bottom row in Figs.~\ref{fig:7-first} and~\ref{fig:13-first}) also
shows a larger difference between BLM-1 and BLM-2 values for $k_J$
in bin-1 and less so for $k_J$ in bin-2. Here, the `red' corrections lower the LLA estimate
whereas the `green' ones make the corresponding LLA estimate less negative.
The corrections are smaller than the ones for $R^{12}_{33}$ and somehow larger 
than the corrections for $R^{12}_{22}$, specifically, `blue' $\sim 5\%$,
`red' $\sim 5\%$ and `green' $\sim 15\%$. Noticeably, while for $R^{12}_{22}$ and
$R^{12}_{33}$ the corrections are very similar at  $\sqrt{s} = 7$ and  $\sqrt{s} = 13$ TeV,
the `green'  $R^{22}_{33}$  receives larger corrections at $\sqrt{s} = 7$ TeV.

One important conclusion we would like to draw 
after comparing Figs.~\ref{fig:7-first} and~\ref{fig:13-first}
is that, in general, for most of the observables there are no
striking changes when we increase the colliding energy from
7 to 13 TeV. This indicates that a sort of asymptotic regime has been approached for the kinematical configurations included in our analysis. It also tells us that our observables
are really as insensitive as possible to effects
which have their origin outside the BFKL dynamics and which 
normally cannot be isolated  ({\it e.g.} influence from the PDFs) with a possible exclusion
at the higher end of the plots, when $Y \sim 8.5-9$. There, some of the observables 
and by that we mean the `red', `green'
or `blue' cases of  $R^{12}_{22}$, $R^{12}_{33}$ and $R^{22}_{33}$,
exhibit a more curved rather than linear behaviour with $Y$ at $\sqrt{s} = 7$ TeV.

\section{$R^{12}_{22}$, $R^{12}_{33}$ and $R^{22}_{33}$ after integration over a central jet rapidity bin}

\begin{figure}[H]
 \hspace{-1.2cm}
 \includegraphics[scale=0.4]{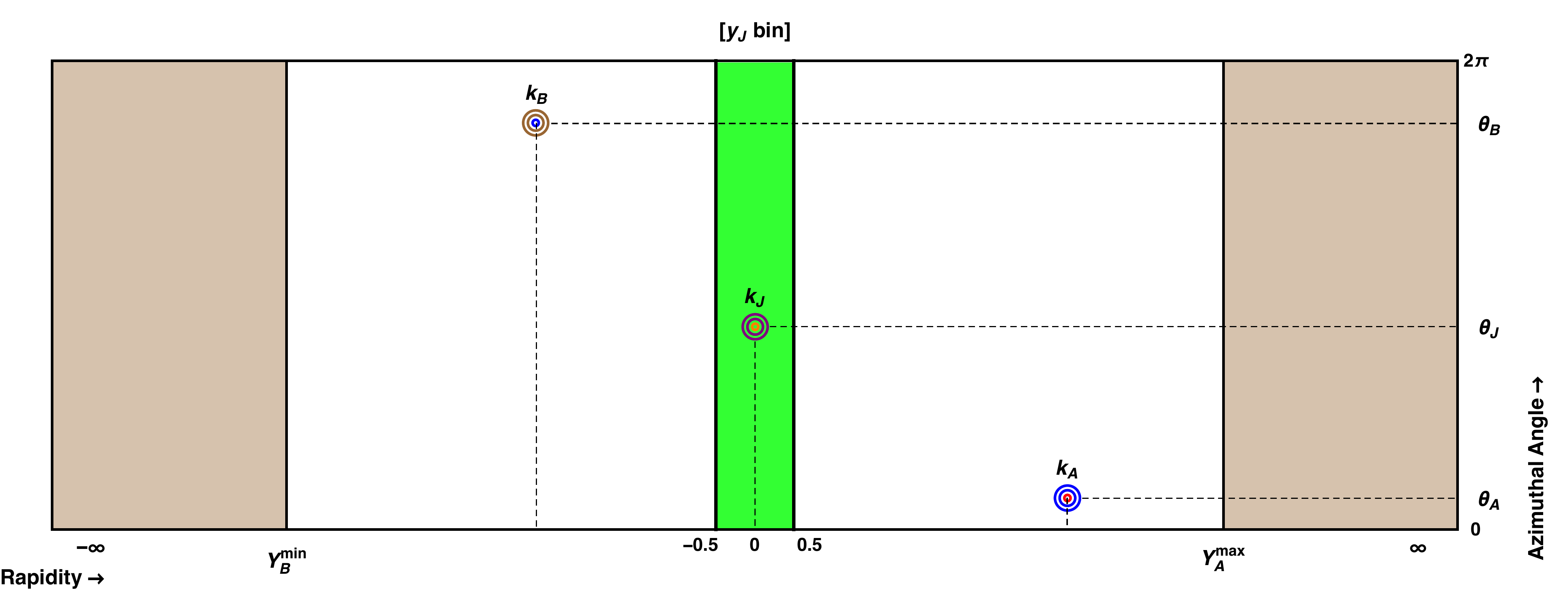}
 \caption[]
 {A primitive lego plot 
 depicting a three-jet event similar to Fig.~\ref{fig:lego2}. Here, however,
 the central jet can take any value in the rapidity range $-0.5 < y_J < 0.5$.
  
 }
 \label{fig:lego2}
 \end{figure}

\begin{figure}[p]
\newgeometry{left=-10cm,right=1cm}
\vspace{-2cm}
\centering

   \hspace{-16.25cm}
   \includegraphics[scale=0.3]{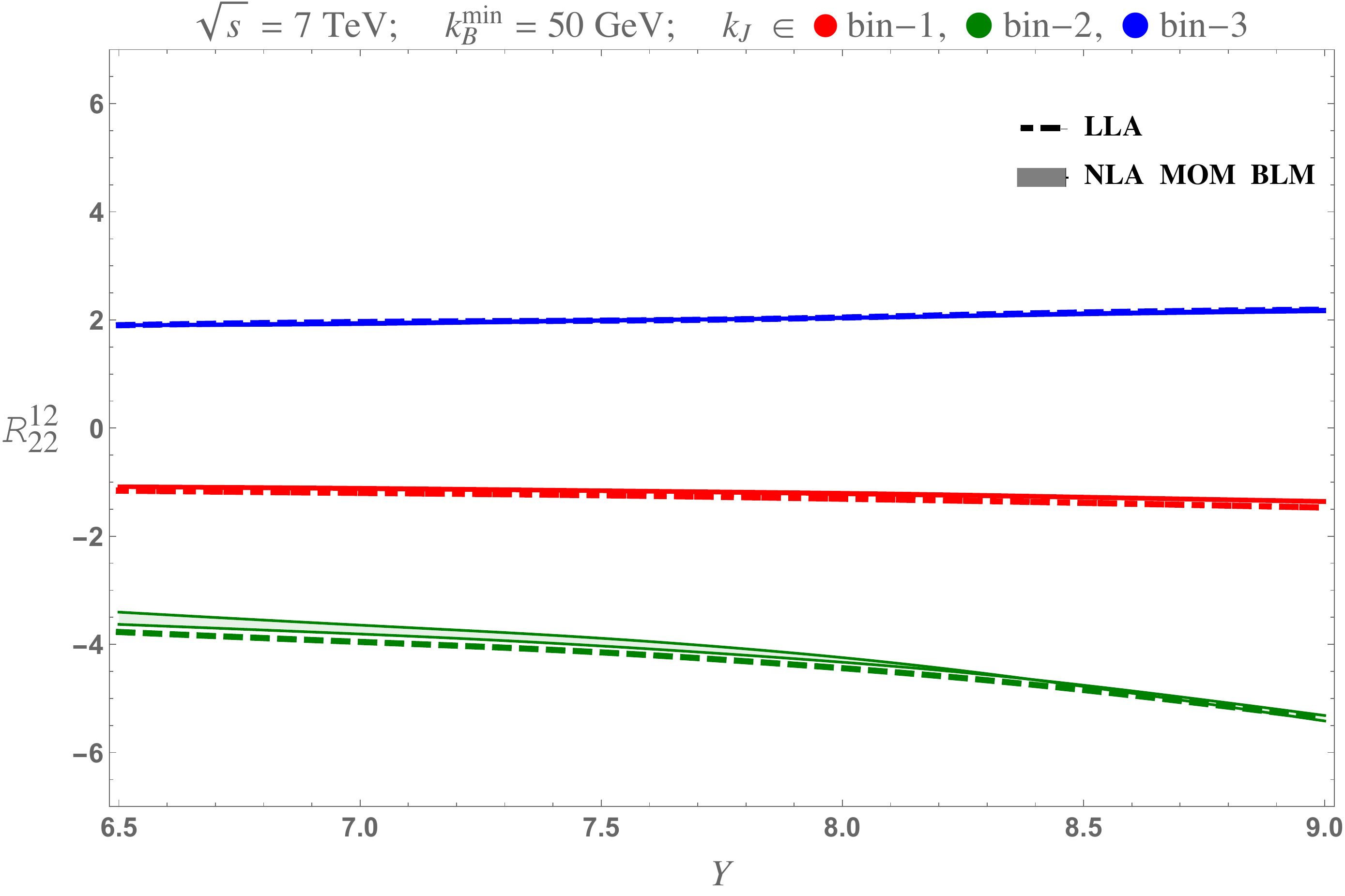}
   \includegraphics[scale=0.3]{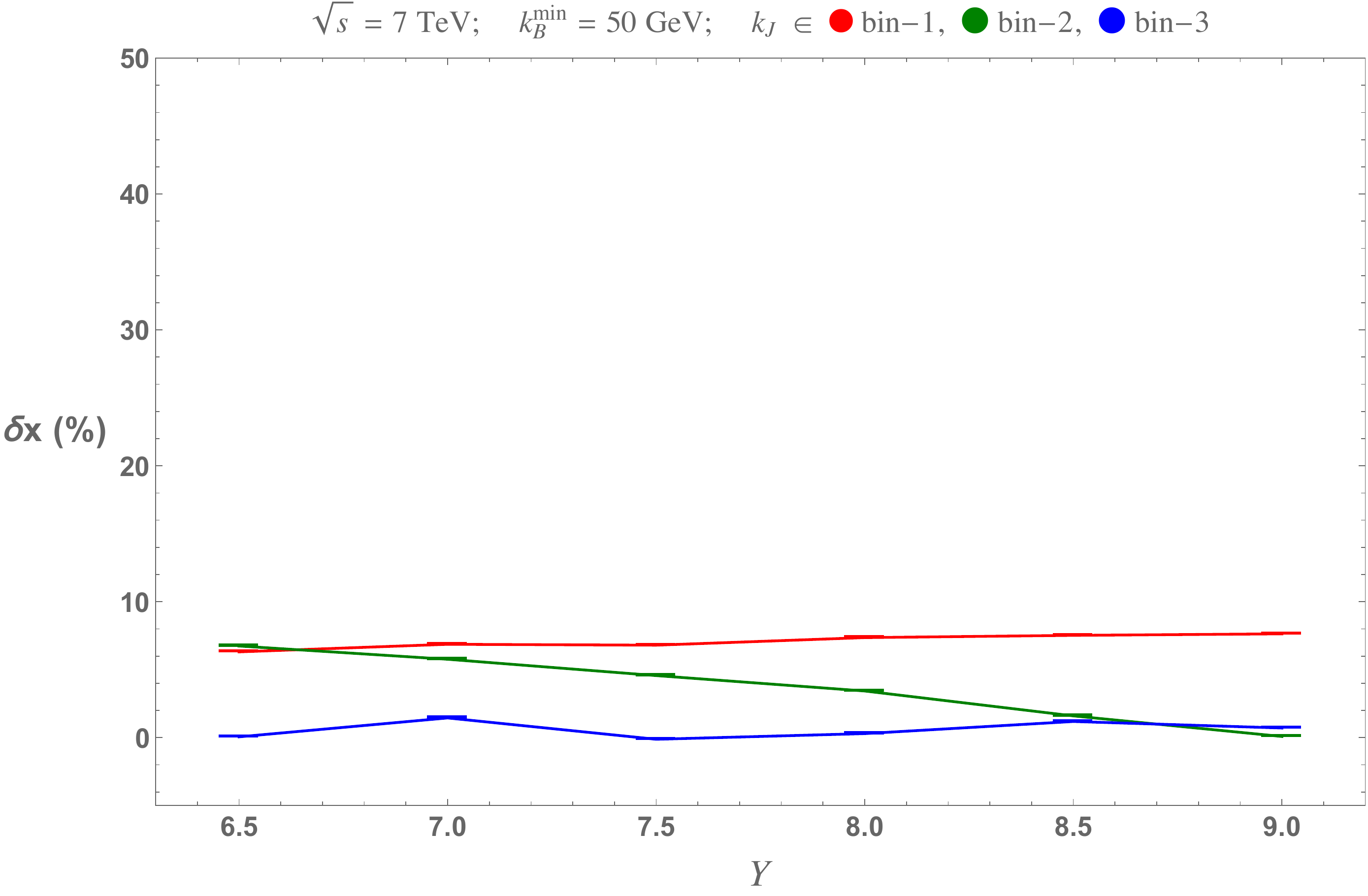}
   \vspace{1cm}

   \hspace{-16.25cm}
   \includegraphics[scale=0.3]{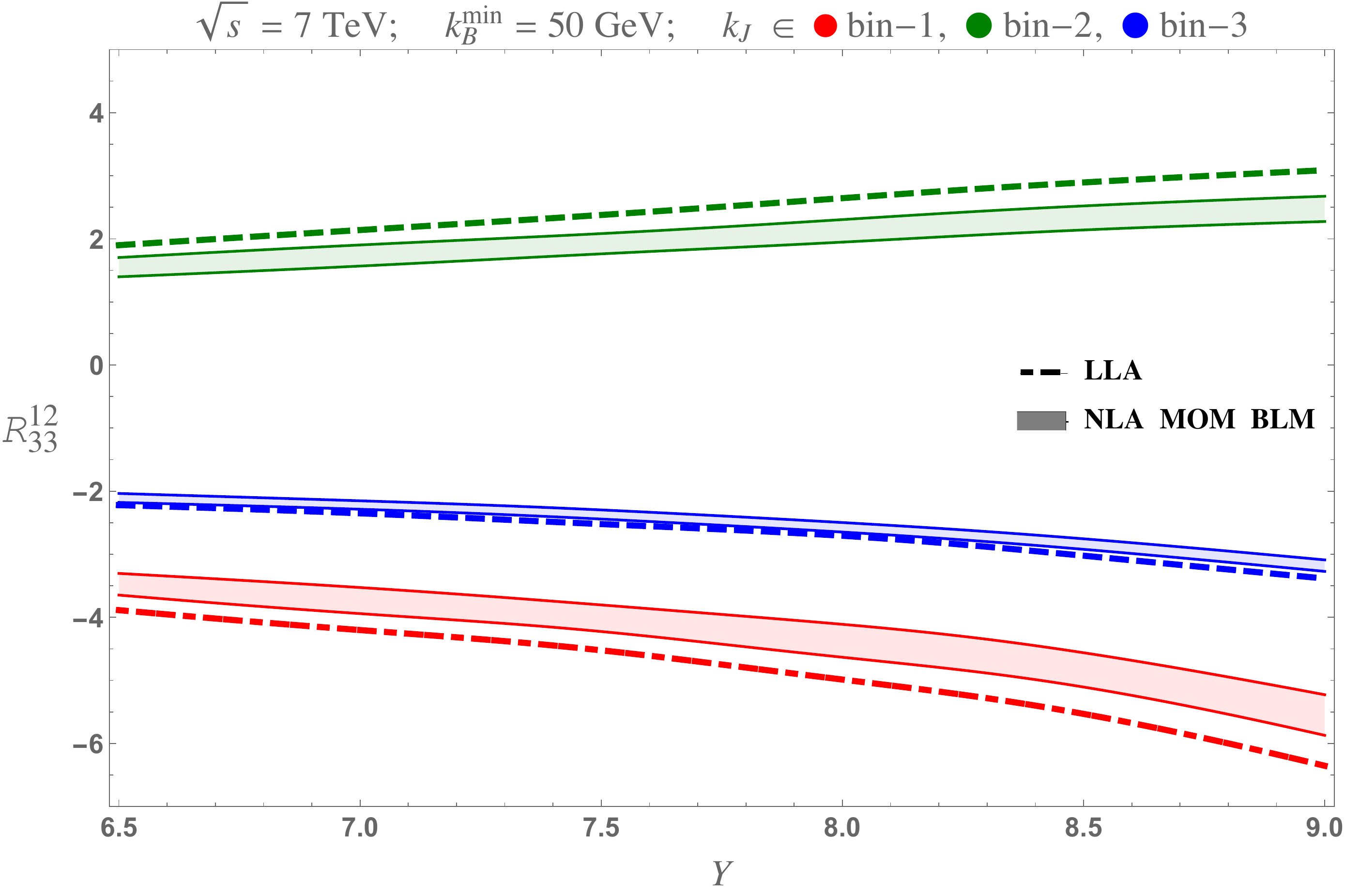}
   \includegraphics[scale=0.3]{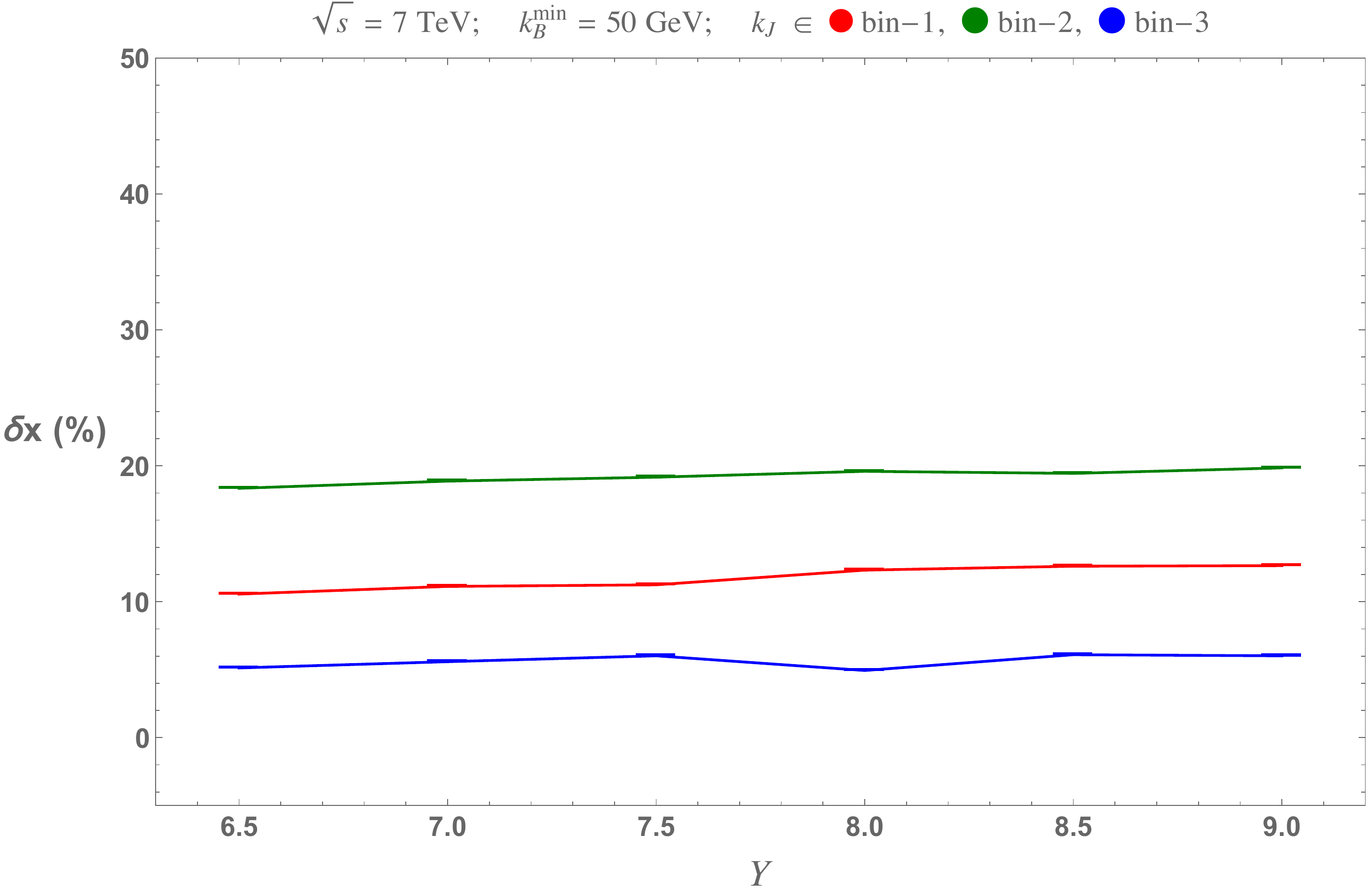}
   \vspace{1cm}

   \hspace{-16.25cm}   
   \includegraphics[scale=0.3]{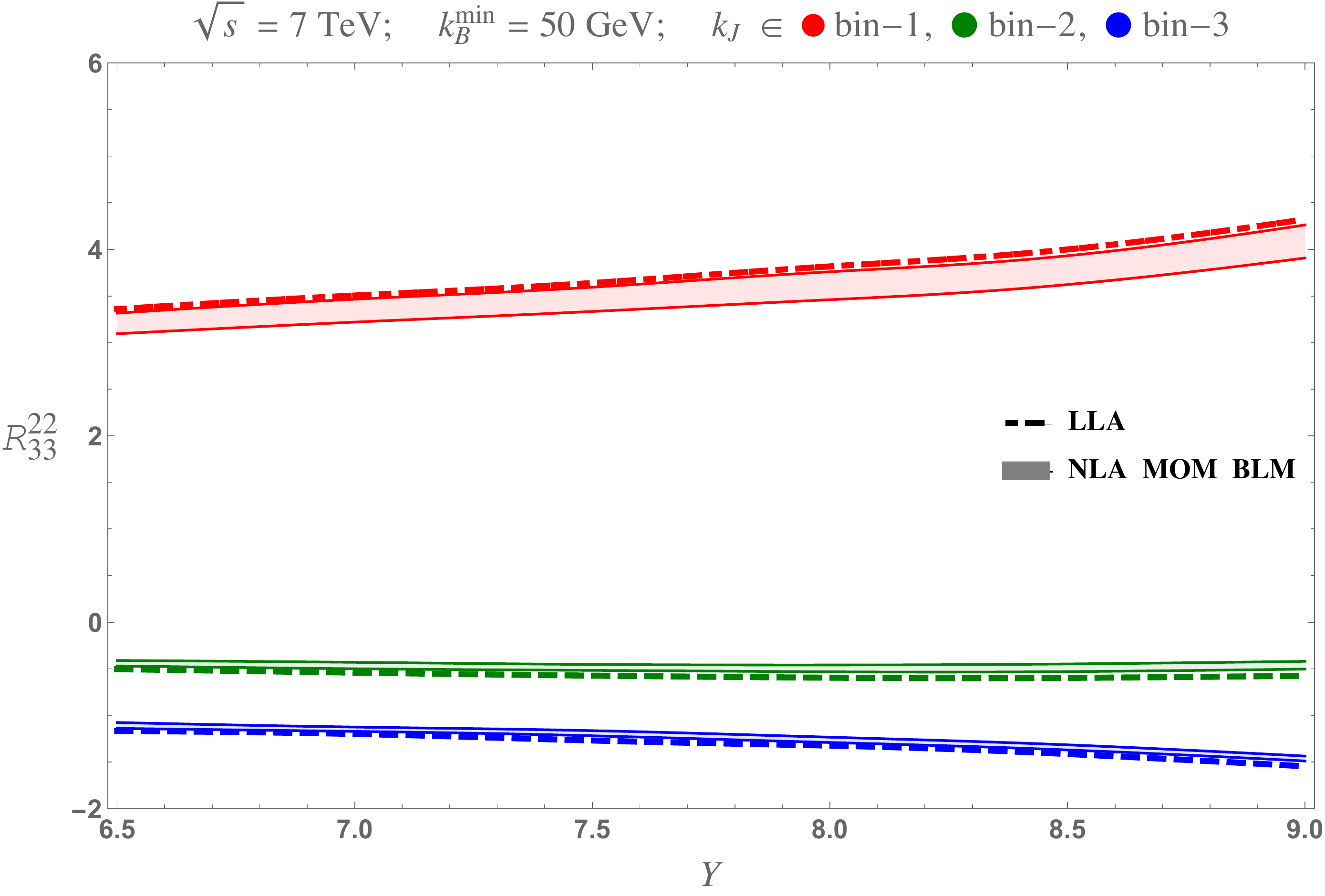}
   \includegraphics[scale=0.3]{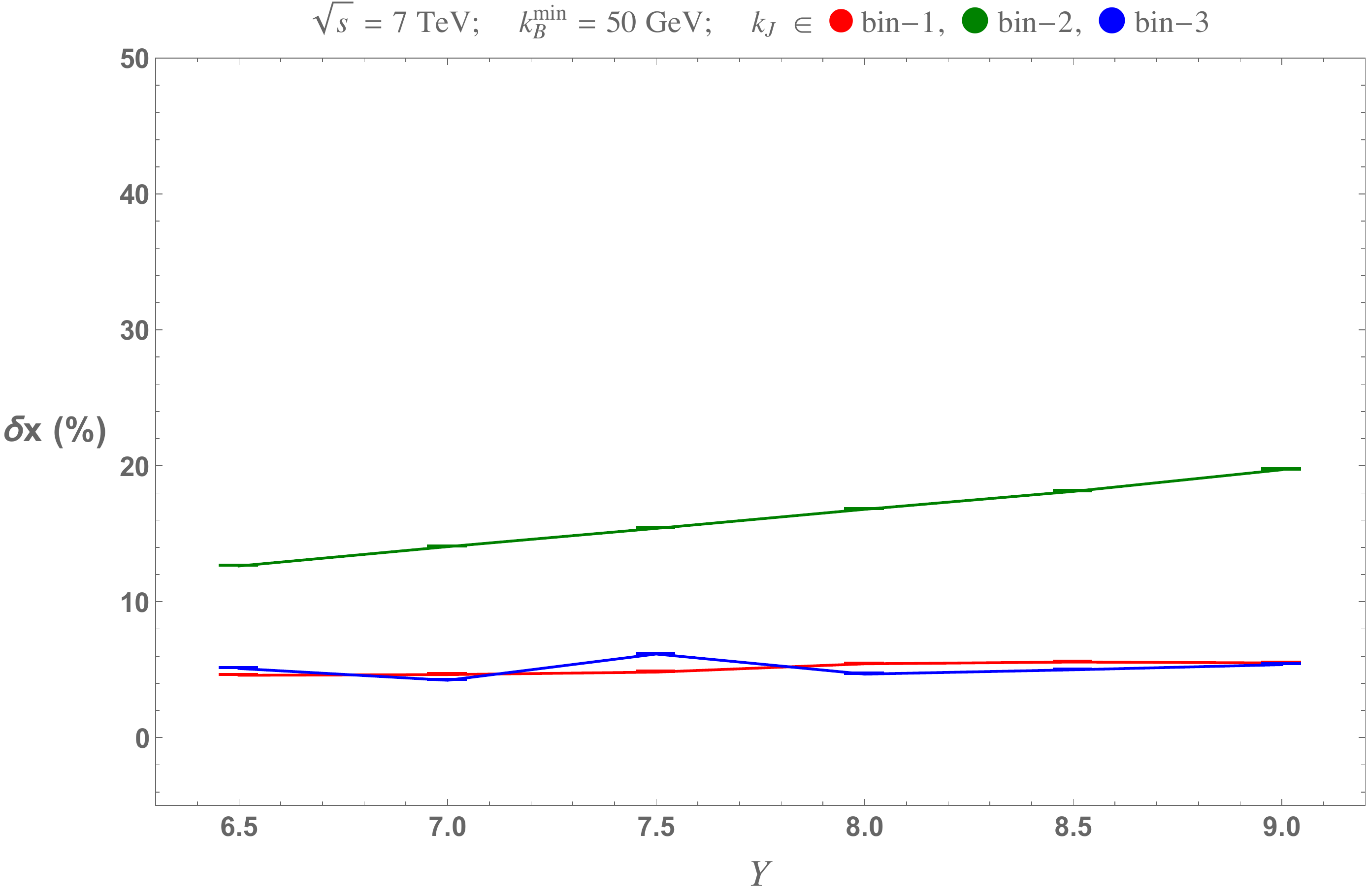}

\restoregeometry
\caption{\small $Y$-dependence of the LLA and NLLA
$R^{12}_{22}$, $R^{12}_{33}$ and $R^{22}_{33}$ at $\sqrt s = 7$ TeV with $y_J$ integrated over a central
rapidity bin
 (left) and the relative NLLA to LLA corrections  (right).} 
\label{fig:7-second}
\end{figure}

\begin{figure}[p]
\newgeometry{left=-10cm,right=1cm}
\vspace{-2cm}
\centering

   \hspace{-16.25cm}
   \includegraphics[scale=0.3]{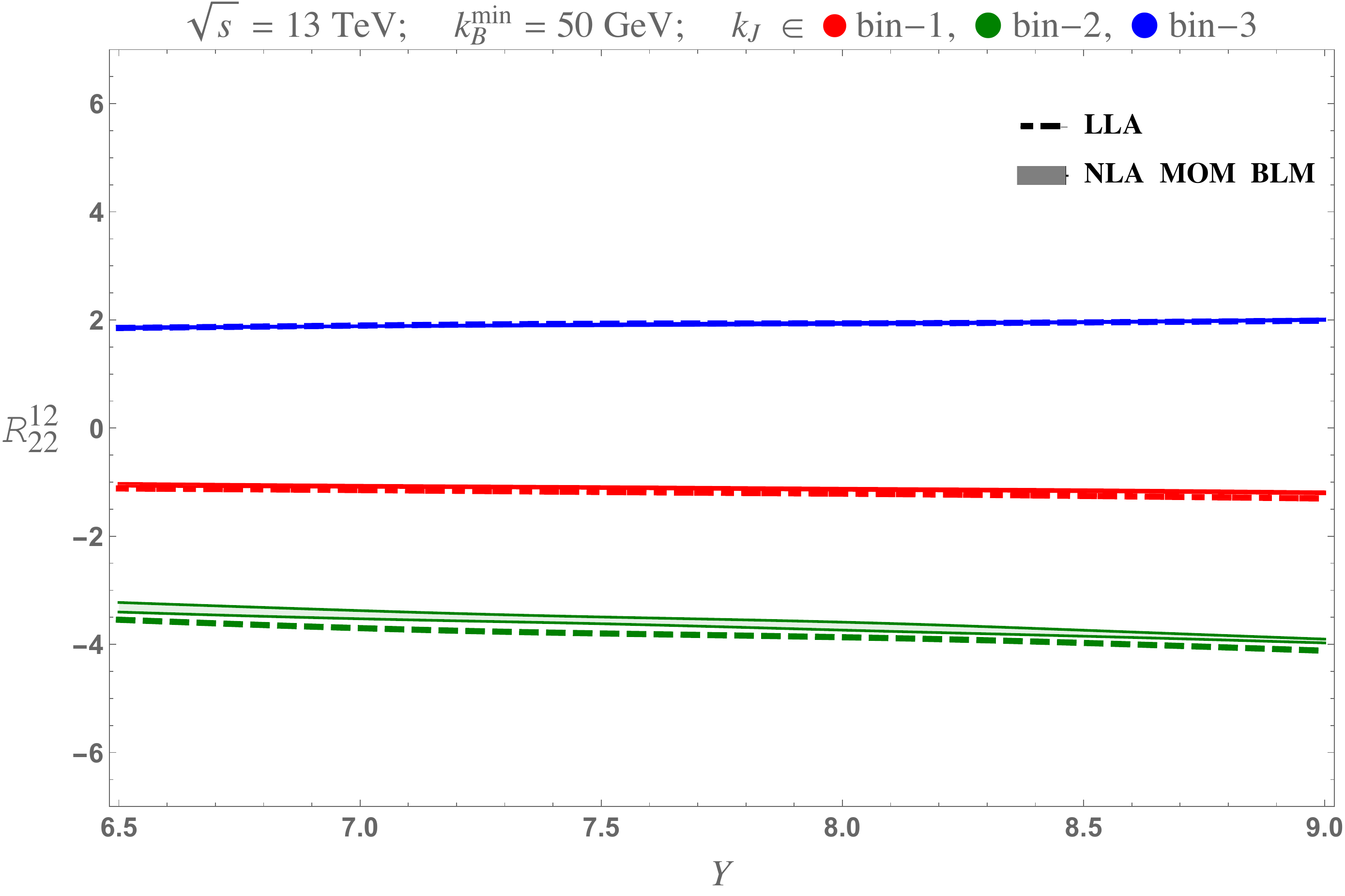}
   \includegraphics[scale=0.3]{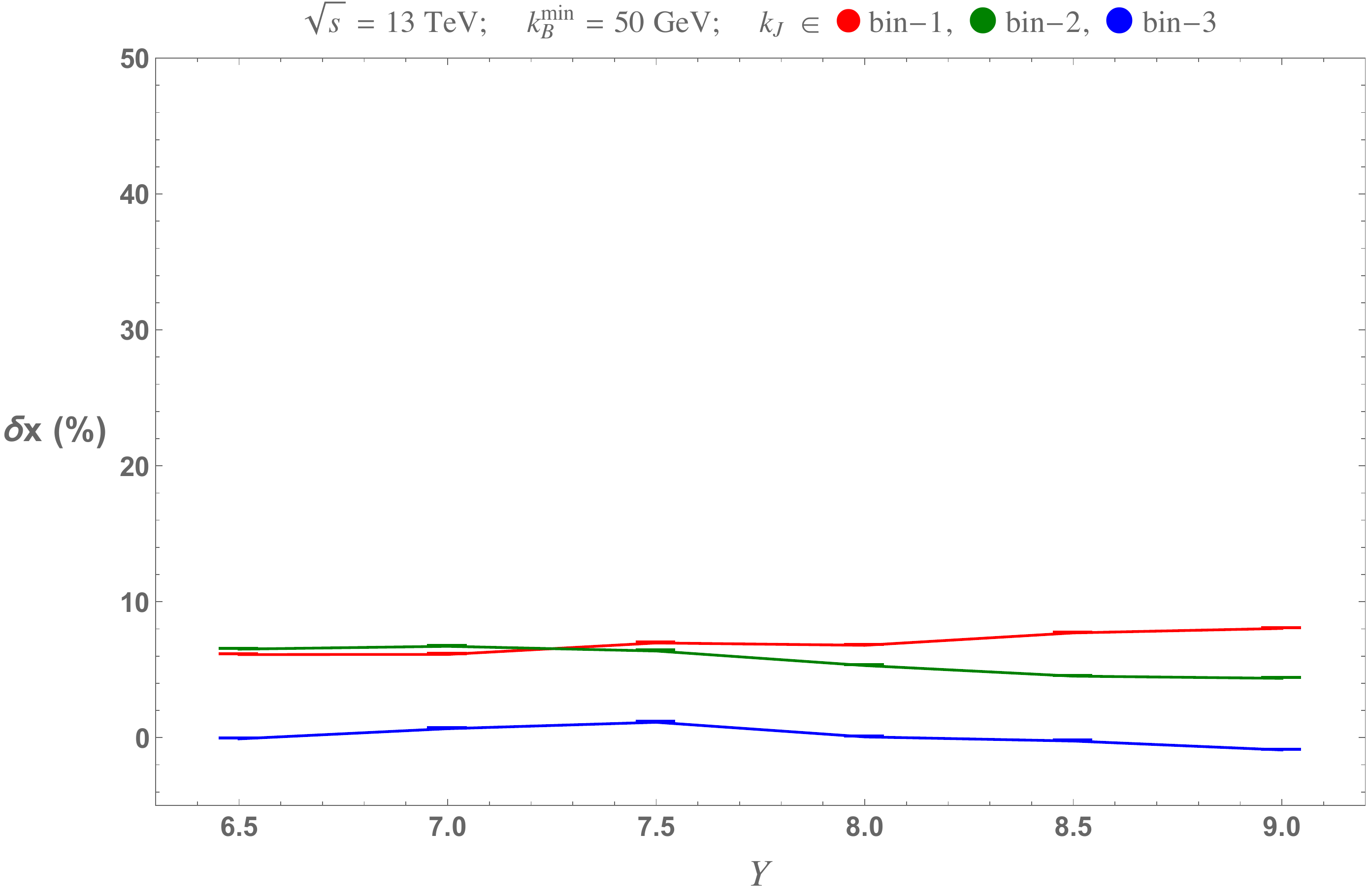}
   \vspace{1cm}

   \hspace{-16.25cm}
   \includegraphics[scale=0.3]{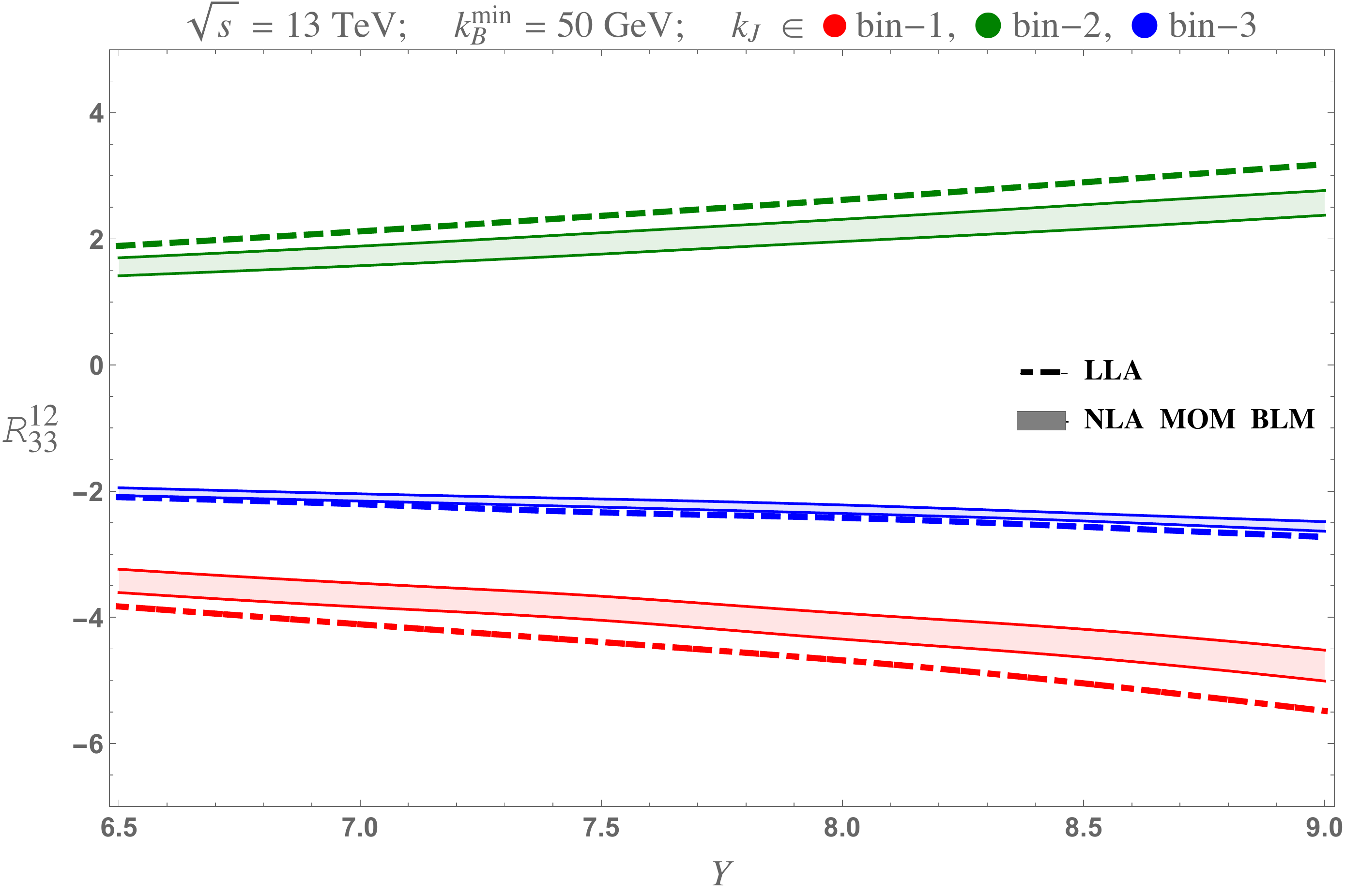}
   \includegraphics[scale=0.3]{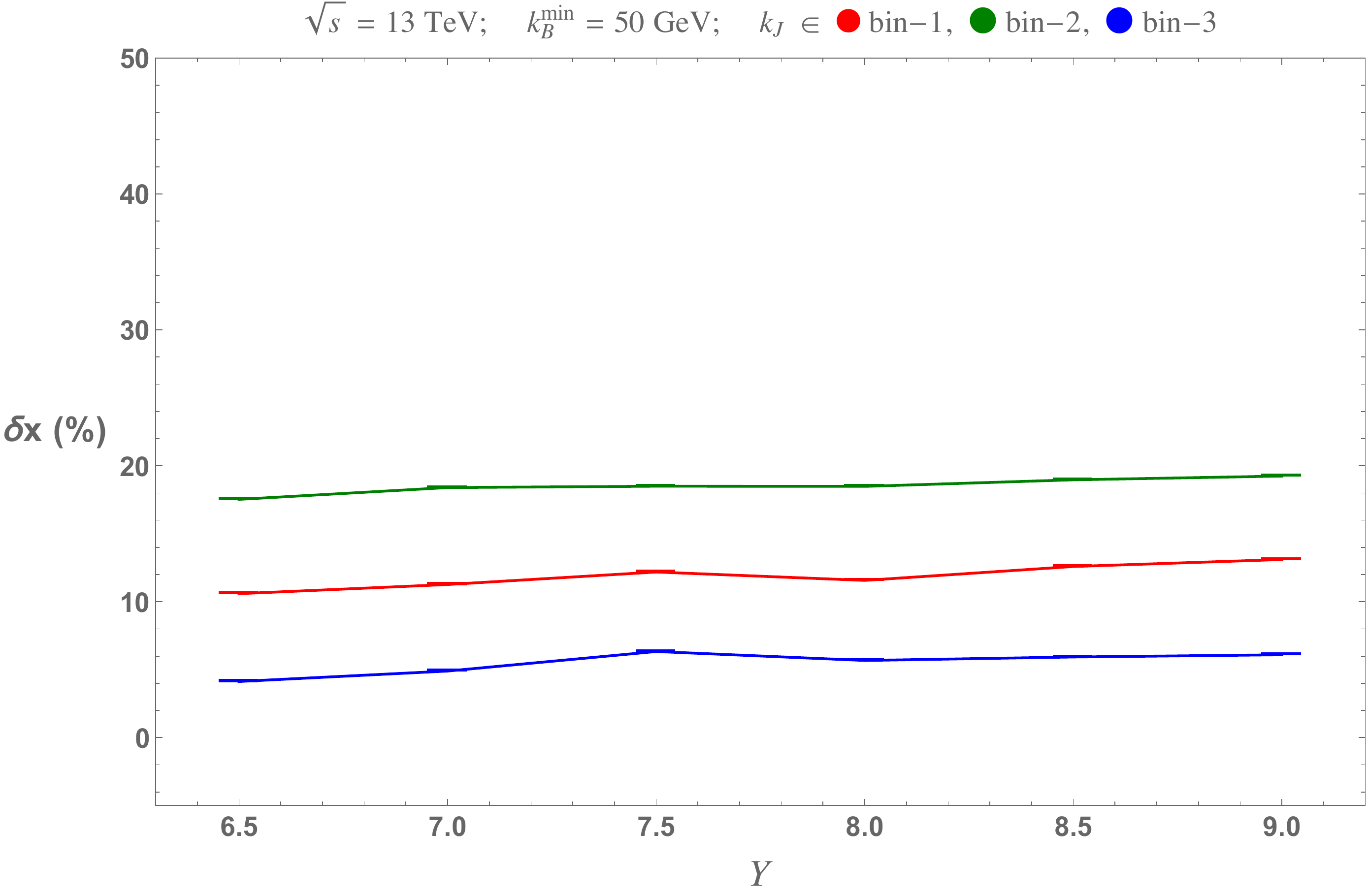}
   \vspace{1cm}

   \hspace{-16.25cm}   
   \includegraphics[scale=0.3]{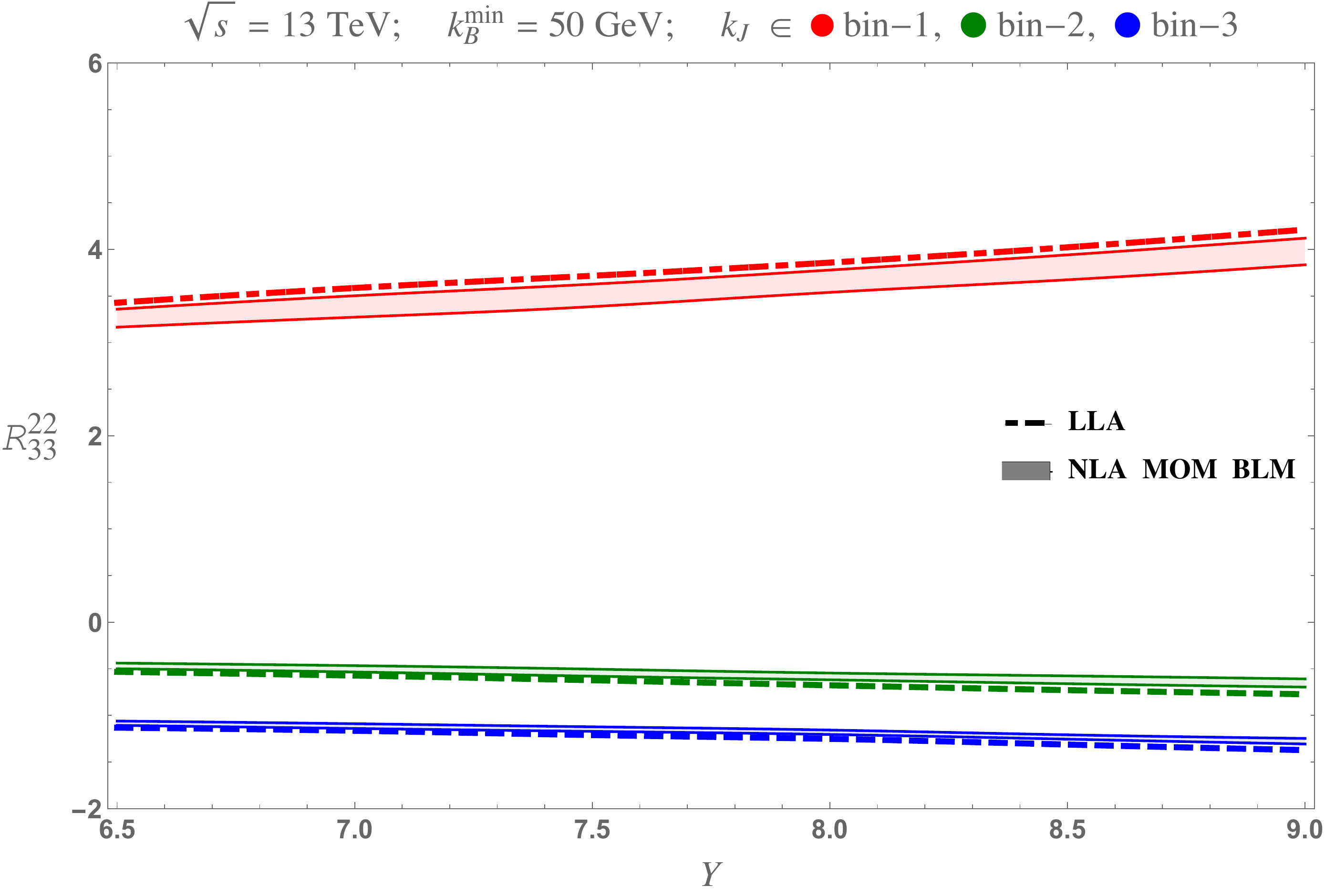}
   \includegraphics[scale=0.3]{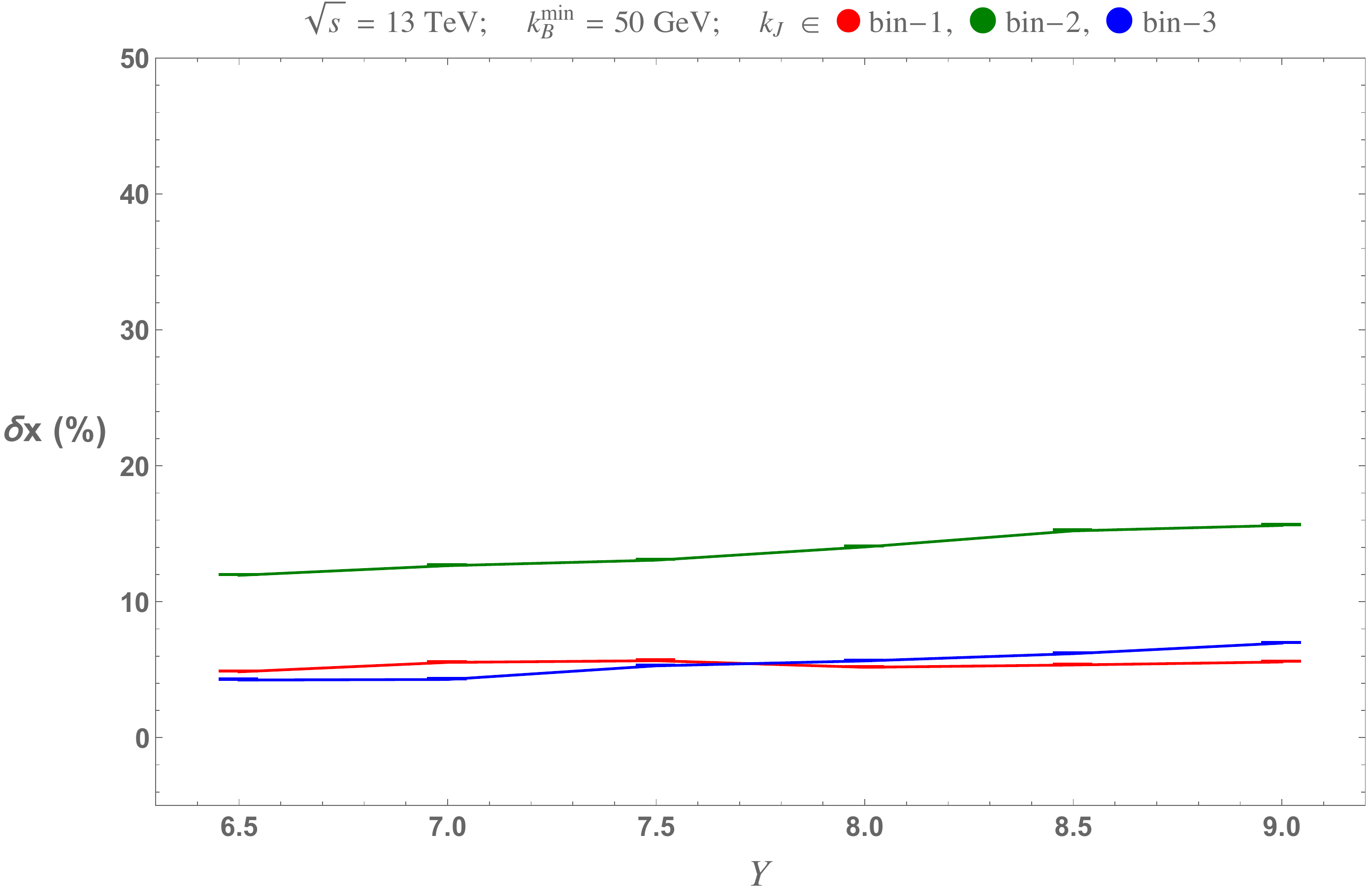}

\restoregeometry
\caption{\small $Y$-dependence of the LLA and NLLA
$R^{12}_{22}$, $R^{12}_{33}$ and $R^{22}_{33}$  at $\sqrt s = 13$ TeV with $y_J$ integrated over a central
rapidity bin (left) and the relative NLLA to LLA corrections  (right).} 
\label{fig:13-second}
\end{figure}

In this section, everything is kept the same as in Section 3 with the exemption
of the allowed values for $y_J$ (see Fig.~\ref{fig:lego2}). 
While in the previous section $y_J = (Y_A+Y_B)/2$,
here $y_J$ is not anymore dependent on the rapidity difference between
the outermost jets, $Y$, and is allowed to take values in a 
rapidity bin around $y_J = 0$.
In particular, $-0.5 < y_J < 0.5$, which in turn means that an additional integration
over $y_J$ needs to be considered in Eq.~\ref{Cmn_int} with $y_J^{\rm min} = -0.5$
and  $y_J^{\rm max} = 0.5$:
\begin{align}
\label{Cmn_int2}
 & 
 C_{MN}^{\text{integ}} =
 \nonumber \\
 &
 \int_{y_J^{\rm min}}^{y_J^{\rm max}} \hspace{-0.25cm} dy_J
 \int_{Y_A^{\rm min}}^{Y_A^{\rm max}} \hspace{-0.25cm} dY_A
 \int_{Y_B^{\rm min}}^{Y_B^{\rm max}} \hspace{-0.25cm} dY_B
 \int_{k_A^{\rm min}}^{k_A^{\rm max}} \hspace{-0.25cm} dk_A
 \int_{k_B^{\rm min}}^{k_B^{\rm max}} \hspace{-0.25cm} dk_B
 \int_{k_J^{\rm min}}^{k_J^{\rm max}} \hspace{-0.25cm} dk_J
 \delta\left(Y_A - Y_B - Y\right) {\cal C}_{MN},
\end{align}

With a slight abuse of notation, we will keep denoting our observables $R_{PQ}^{MN}$:
\begin{eqnarray}
\label{RPQMN}
R_{PQ}^{MN} \, = \, \frac{C_{MN}^{\text{integ}}}{C_{PQ}^{\text{integ}}}\,.
\label{RmnqpNew2}
\end{eqnarray}
Therefore, in Figs.~\ref{fig:7-second} and~\ref{fig:13-second}
we still have $R^{12}_{22}$, $R^{12}_{33}$ and $R^{22}_{33}$ although here
they do contain the extra integration over $y_J$.

We notice immediately that Fig.~\ref{fig:7-first} is very similar to the 
integrated over $y_J$ observables in Fig.~\ref{fig:7-second} and
the same holds for Figs.~\ref{fig:13-first} and~\ref{fig:13-second}. Therefore,
we will not discuss here the individual 
behaviours of $R^{12}_{22}$, $R^{12}_{33}$ and $R^{22}_{33}$
with $Y$, neither the $\delta x(\%)$ corrections, since this would only mean to repeat
the discussion of the previous section.
We would like only to note that the striking similarity between Fig.~\ref{fig:7-first}
and Fig.~\ref{fig:7-second} and between Fig.~\ref{fig:13-first}
and Fig.~\ref{fig:13-second} was to be expected if we remember that the 
partonic-level quantities $\mathcal{R}_{PQ}^{MN}$ do not change noticeably
if we vary the position in rapidity of the central jet, as long as the position remains
``sufficiently" central (see Ref.~\cite{Caporale:2015vya}). 
This property is very important and we will discuss it more in the next section.
Here, we should stress that the observables as presented in this section
can be readily compared to experimental data.

\section{$R^{12}_{22}$, $R^{12}_{33}$ and $R^{22}_{33}$ after integration over a forward, backward and central rapidity bin}

\begin{figure}[H]
 \hspace{-1.2cm}
 \includegraphics[scale=0.4]{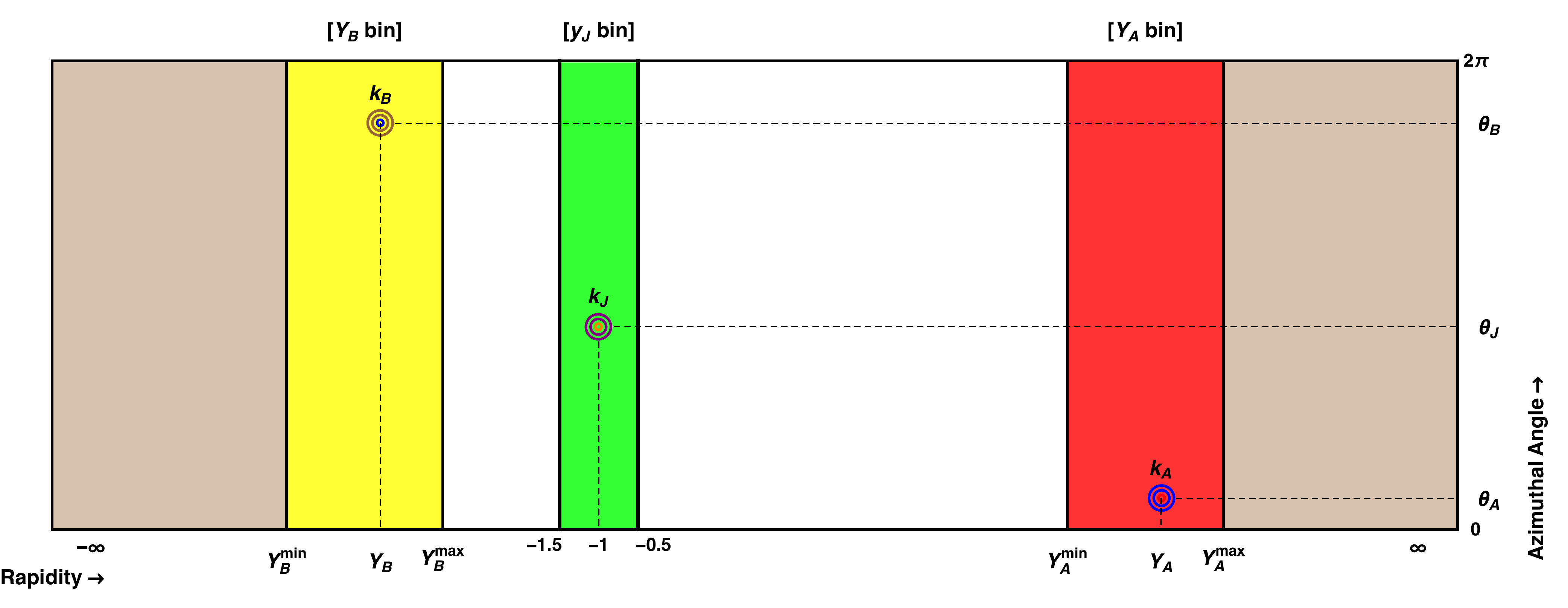}
 \caption[]
 {A primitive lego plot 
 depicting a three-jet event similar to Fig.~\ref{fig:lego2}. Here, however,
 the rapidity of the central jet can take any value in the distinct ranges $y_i-0.5 < y_J < y_i+0.5$, where
 $y_i$ is the central value of the rapidity bin with $y_i = \{-1, -0.5, 0, 0.5, 1\}$.
 In this figure, $y_i = -1$.
 Moreover, $Y = Y_A - Y_B$ is not anymore fixed. Instead, the forward jet has a rapidity restricted in
 the red bin whereas the backward jet in the yellow bin. 
 }
 \label{fig:lego3}
 \end{figure}

In this section, we present an alternative kinematical configuration (see Fig.~\ref{fig:lego3}) for the 
generalised ratios $R_{PQ}^{MN}$. We do this for two reasons.
Firstly, to offer a different setup for which the comparison between theoretical
predictions and experimental data might be easier, compared to the previous
section. Secondly, to demonstrate that the generalised ratios do capture
the Bethe-Salpeter characteristics of the BFKL radiation. The latter needs
a detailed explanation. 

Let us assume that we have a gluonic ladder exchanged 
in the $t$-channel between a
forward jet (at rapidity $Y_A$) and a backward jet (at rapidity $Y_B$)  accounting for minijet activity
between the two jets. By gluonic ladder here we mean the gluon Green function 
$\varphi \left(\vec{p}_A,\vec{p}_B,Y_A - Y_B\right) $, where
$\vec{p}_A$ and $\vec{p}_B$ are the reggeized momenta connected to the forward
and backward jet vertex respectively. 
It is known that the following relation holds for the gluon Green function:
\begin{eqnarray}
\hspace{-.3cm}
\varphi \left(\vec{p}_A,\vec{p}_B,Y_A - Y_B\right) &=& 
\int d^2 \vec{k} \, 
\varphi \left(\vec{p}_A,\vec{k},Y_A - y\right) 
 \varphi \left(\vec{k},\vec{p}_B,y - Y_B\right).
 \label{BasicRelation0}
\end{eqnarray}
In other words, one may `cut' the gluonic ladder at any rapidity $y$ between 
$Y_A$ and $Y_B$ and then integrate
over the reggeized momentum $\vec{k}$ that flows
in the $t$-channel, to recover the initial ladder.
Which value of $y$ one chooses to `cut' the ladder at is irrelevant.
Therefore, observables directly connected to a realisation of the r.h.s of Eq.~\ref{BasicRelation0}
should display this $y$-independence.

In our study actually, we have a very similar picture as the one described in the r.h.s of Eq.~\ref{BasicRelation0}. The additional element is that we do not only `cut' the gluonic ladder but
we also `insert' a jet vertex for the central jet. This means that the $y$-independence we discussed
above should be present in one form or another. 
To be precise, we do see the $y$-independence behaviour
but now we have to consider the additional constraint that $y$ cannot take any extreme values,
that is, it cannot be close to $Y_A$ or $Y_B$. For a more detailed discussion
of Eq.~\ref{BasicRelation0}, we refer the
reader to Appendix A, here we will proceed to present our numerical results.

The kinematic setup now is different than in the previous sections. We allow $Y_A$ and $Y_B$
to take
values such that $(Y_A^{\text{min}} = 3) < Y_A < (Y_A^{\text{max}} = 4.7)$ and
$(Y_B^{\text{min}} = -4.7) < Y_B < (Y_B^{\text{max}} = -3)$. Moreover, we allow for
the rapidity of the central jet to take values in five distinct rapidity bins of unit width, that is,
$y_i-0.5 < y_J<y_i+0.5$, with $y_i = \{-1, -0.5, 0, 0.5, 1\}$ and we define the
coefficients $C_{MN}^{\rm integ}(y_i)$ as function of $y_i$:
\begin{align}
\label{Cmn_int3}
 & 
 C_{MN}^{\rm integ}(y_i) =
 \nonumber \\
 &
  \int_{y_i-0.5}^{y_i+0.5} \hspace{-0.25cm} dy_J
 \int_{Y_A^{\rm min}}^{Y_A^{\rm max}} \hspace{-0.25cm} dY_A
 \int_{Y_B^{\rm min}}^{Y_B^{\rm max}} \hspace{-0.25cm} dY_B
 \int_{k_A^{\rm min}}^{k_A^{\rm max}} \hspace{-0.25cm} dk_A
 \int_{k_B^{\rm min}}^{k_B^{\rm max}} \hspace{-0.25cm} dk_B
 \int_{k_J^{\rm min}}^{k_J^{\rm max}} \hspace{-0.25cm} dk_J\,\,
{\mathcal C}_{MN}.
\end{align}

\begin{figure}[p]
\newgeometry{left=-10cm,right=1cm}
\vspace{-2cm}
\centering

   \hspace{-16.25cm}
   \includegraphics[scale=0.3]{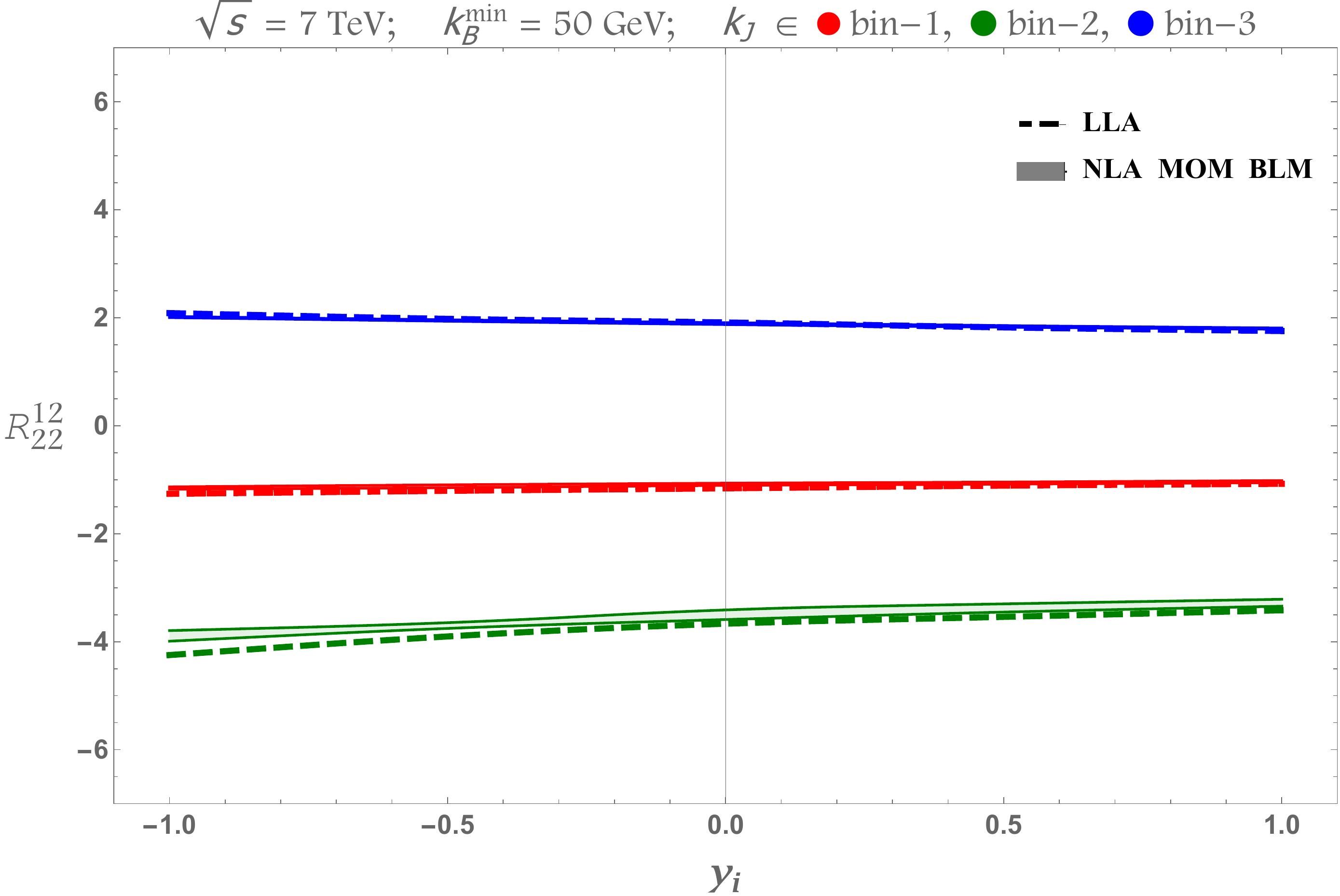}
   \includegraphics[scale=0.3]{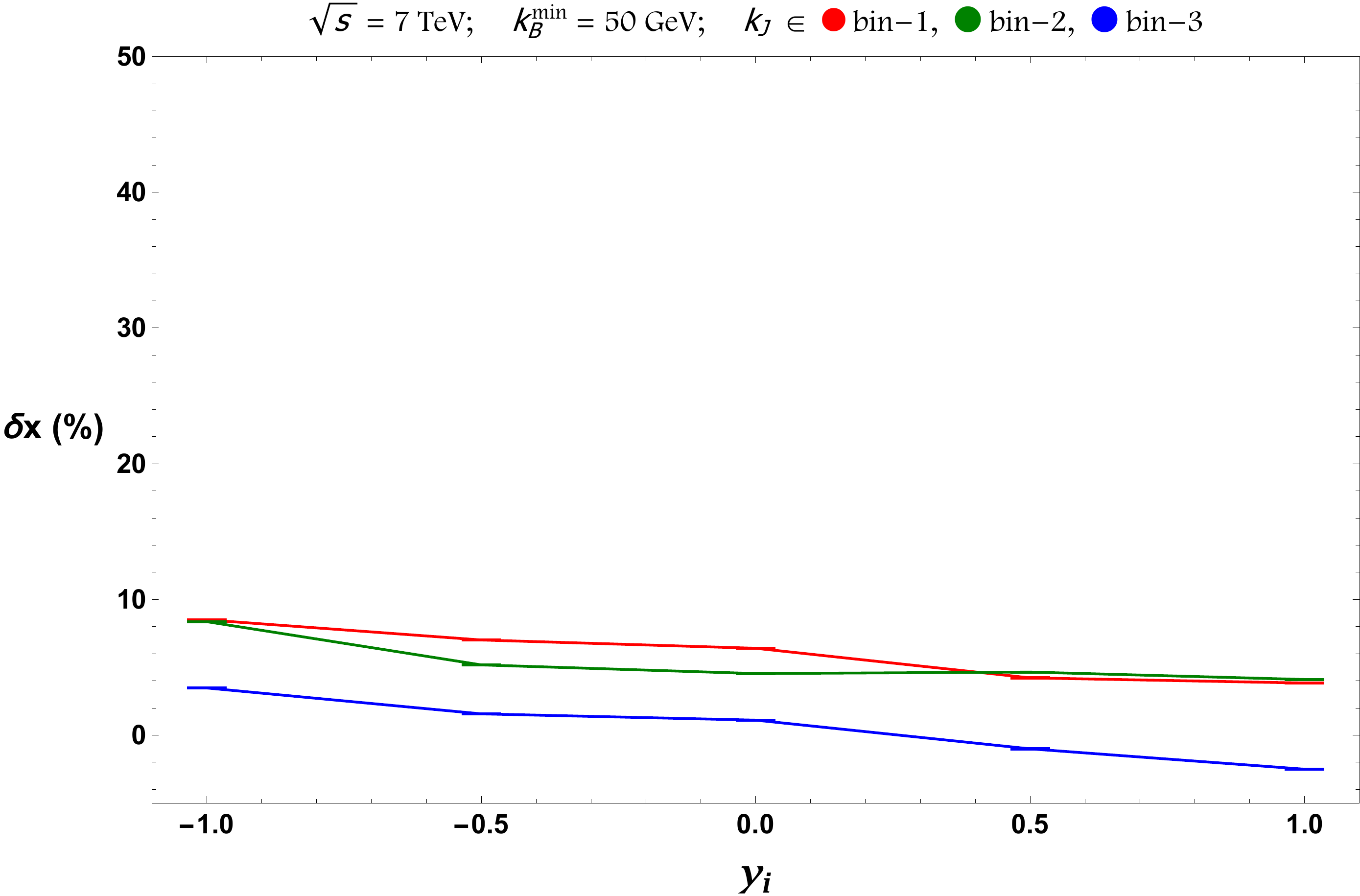}
   \vspace{1cm}

   \hspace{-16.25cm}
   \includegraphics[scale=0.3]{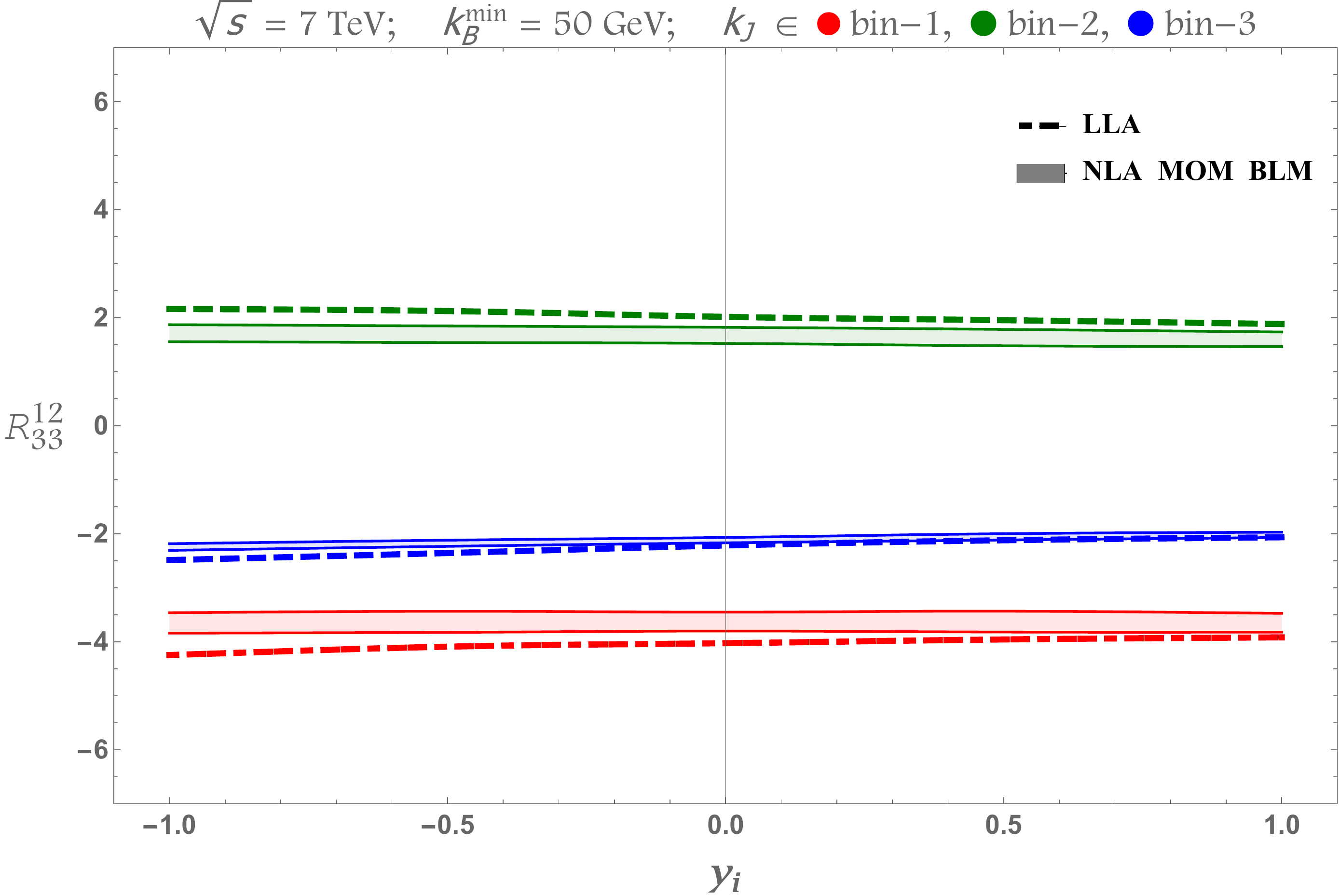}
   \includegraphics[scale=0.3]{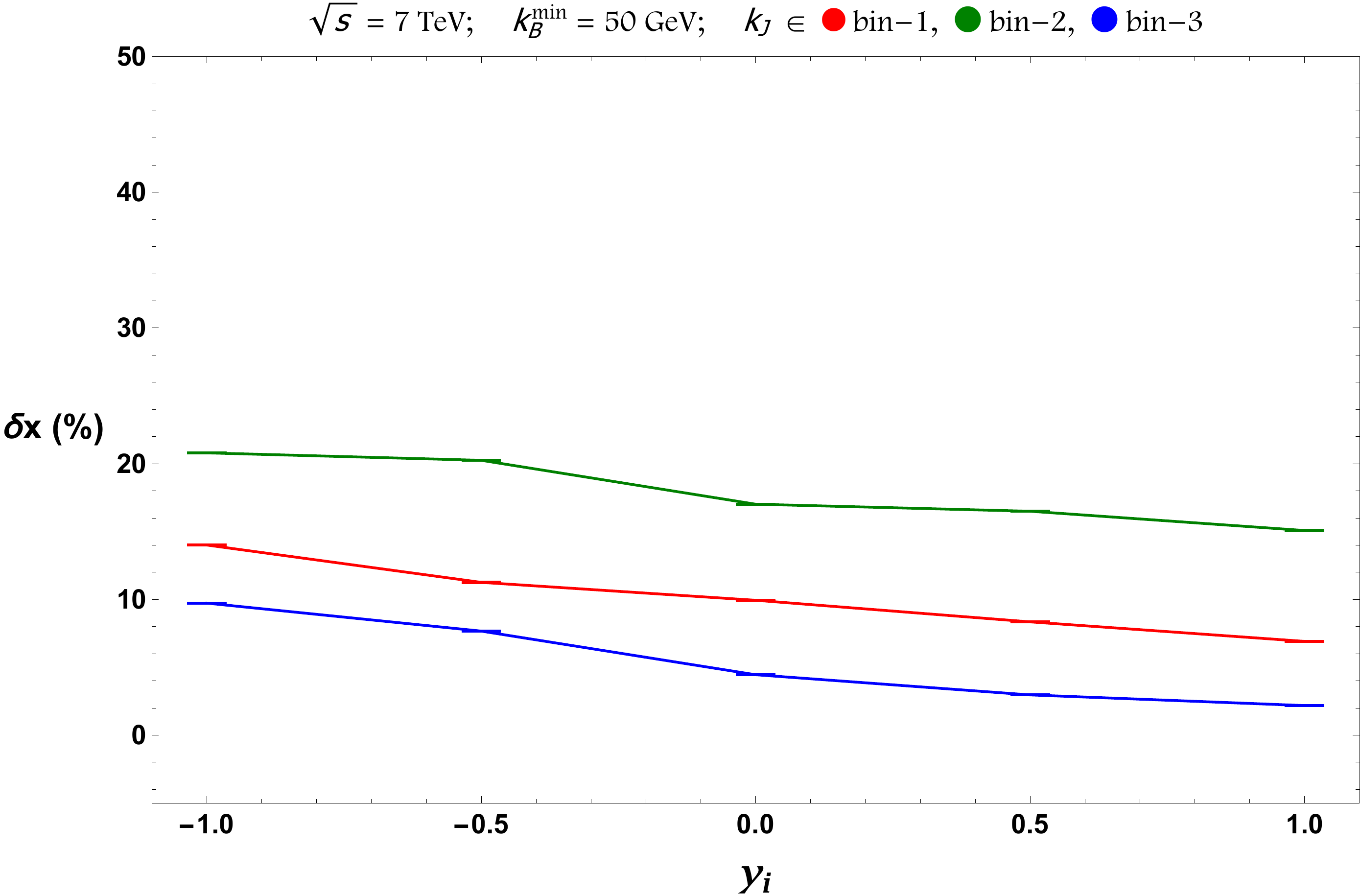}
   \vspace{1cm}

   \hspace{-16.25cm}   
   \includegraphics[scale=0.3]{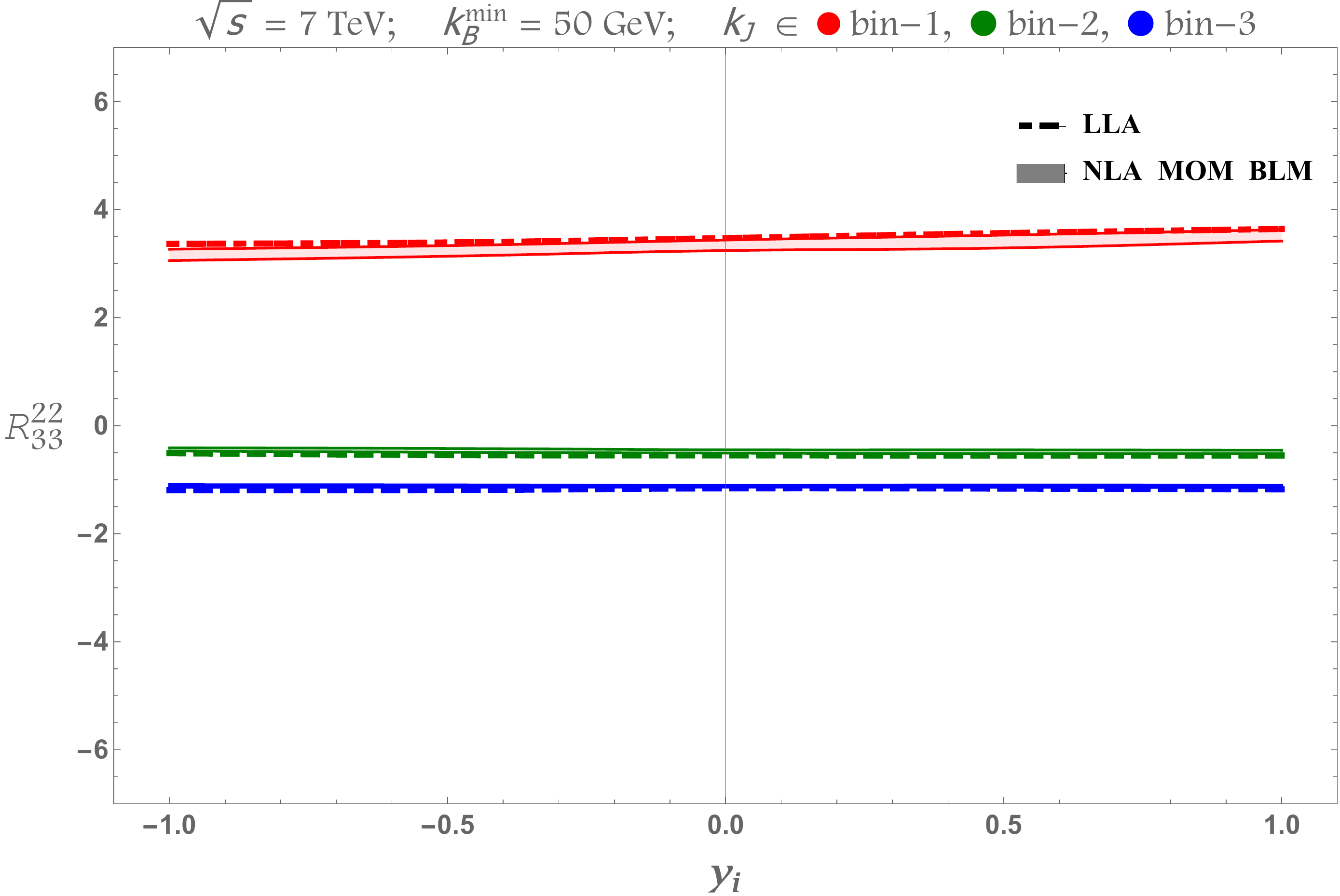}
   \includegraphics[scale=0.3]{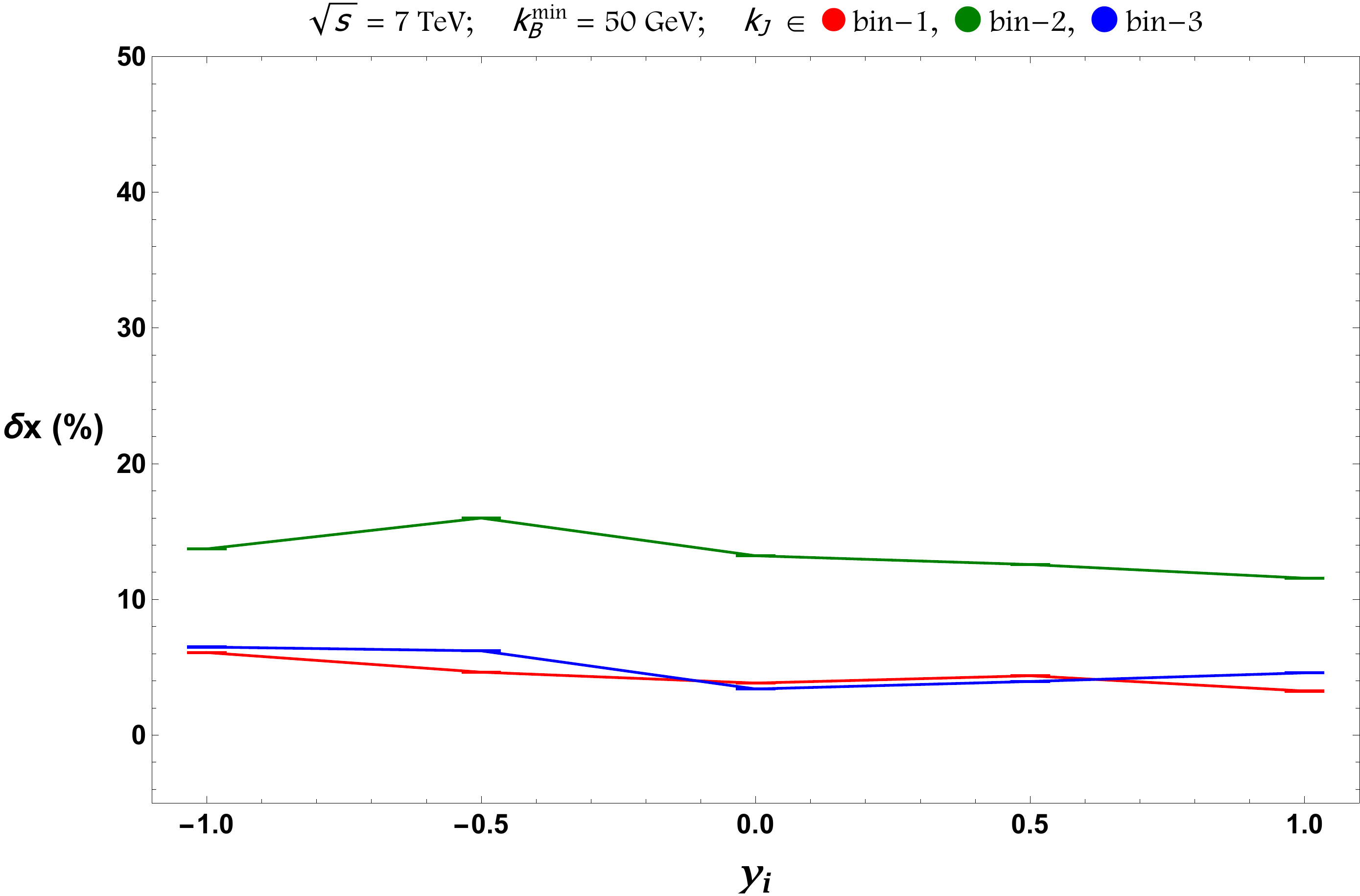}

\restoregeometry
\caption{\small $y_i$-dependence of the LLA and NLLA
$R^{12}_{22}$, $R^{12}_{33}$ and $R^{22}_{33}$ 
at $\sqrt s = 7$ TeV (left) and the relative NLLA to LLA corrections  (right).} 
\label{fig:7-third}
\end{figure}

\begin{figure}[p]
\newgeometry{left=-10cm,right=1cm}
\vspace{-2cm}
\centering

   \hspace{-16.25cm}
   \includegraphics[scale=0.3]{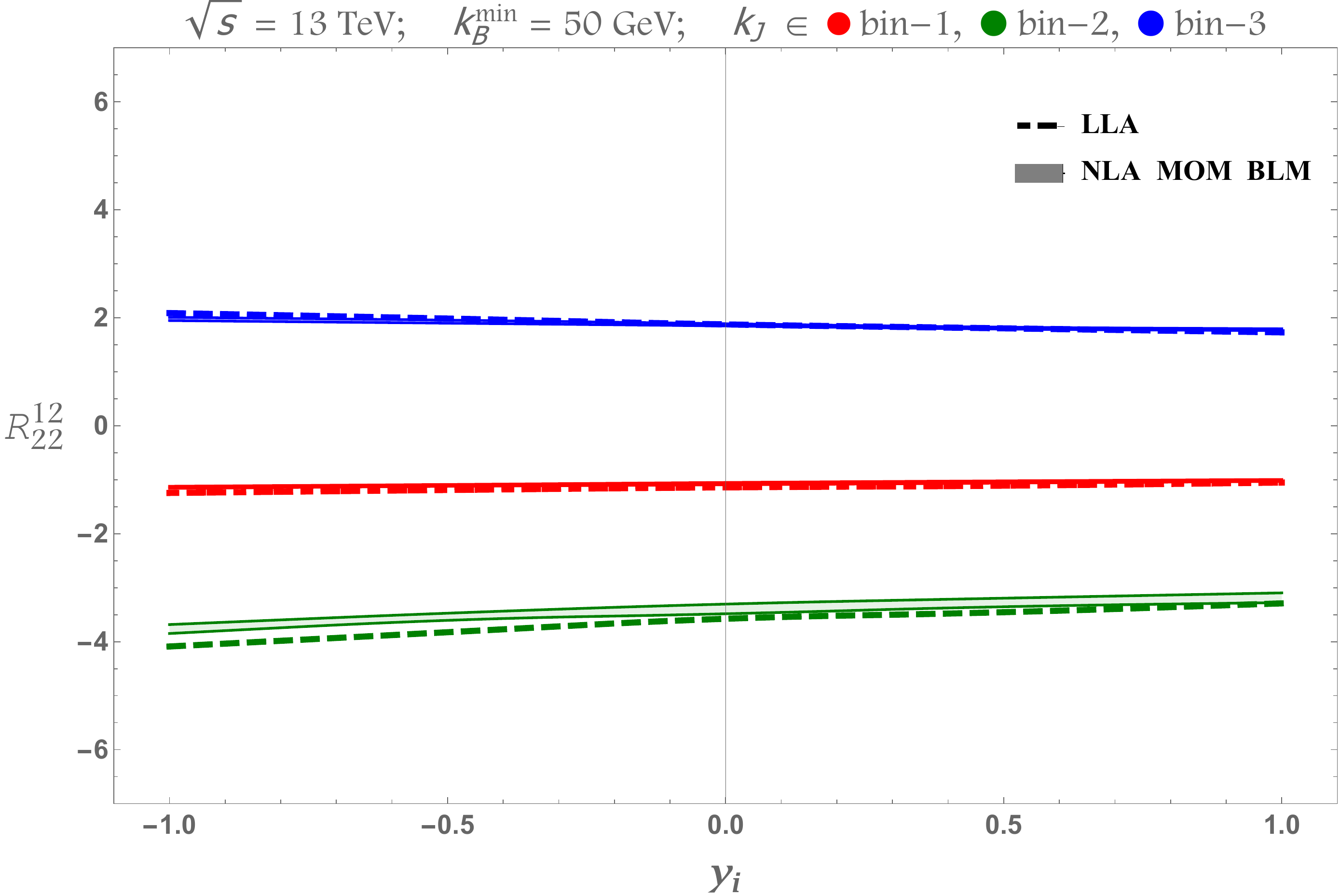}
   \includegraphics[scale=0.3]{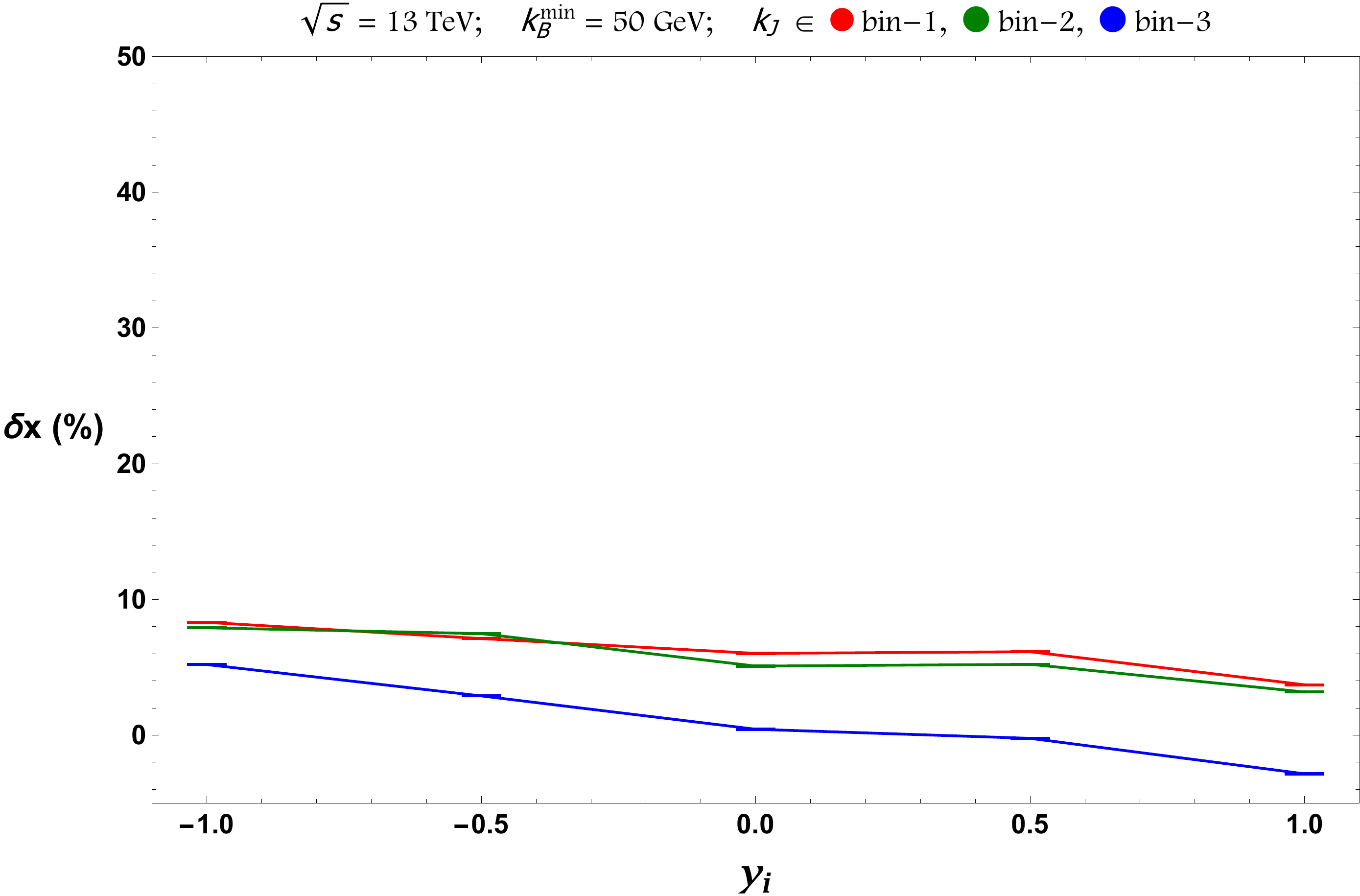}
   \vspace{1cm}

   \hspace{-16.25cm}
   \includegraphics[scale=0.3]{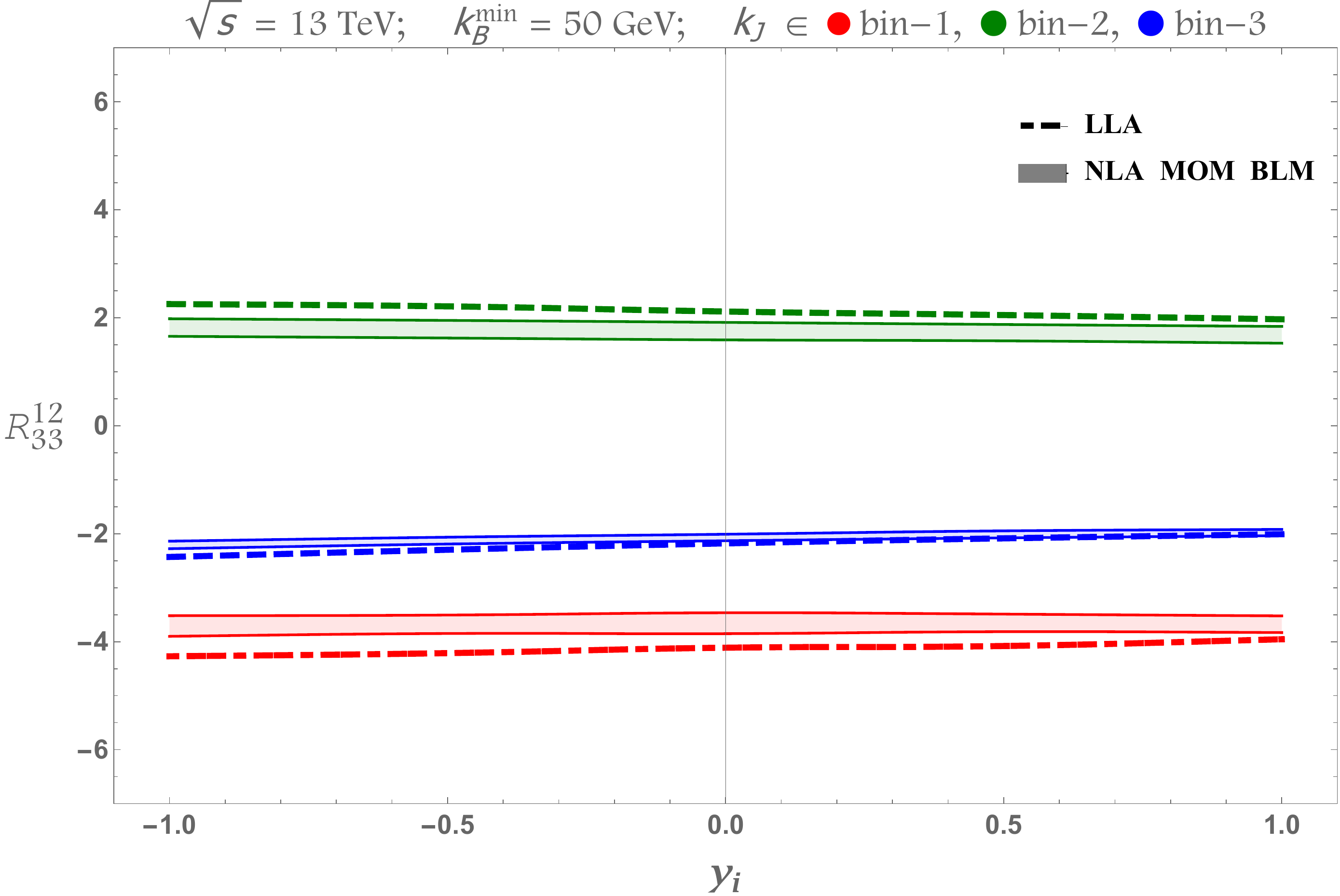}
   \includegraphics[scale=0.3]{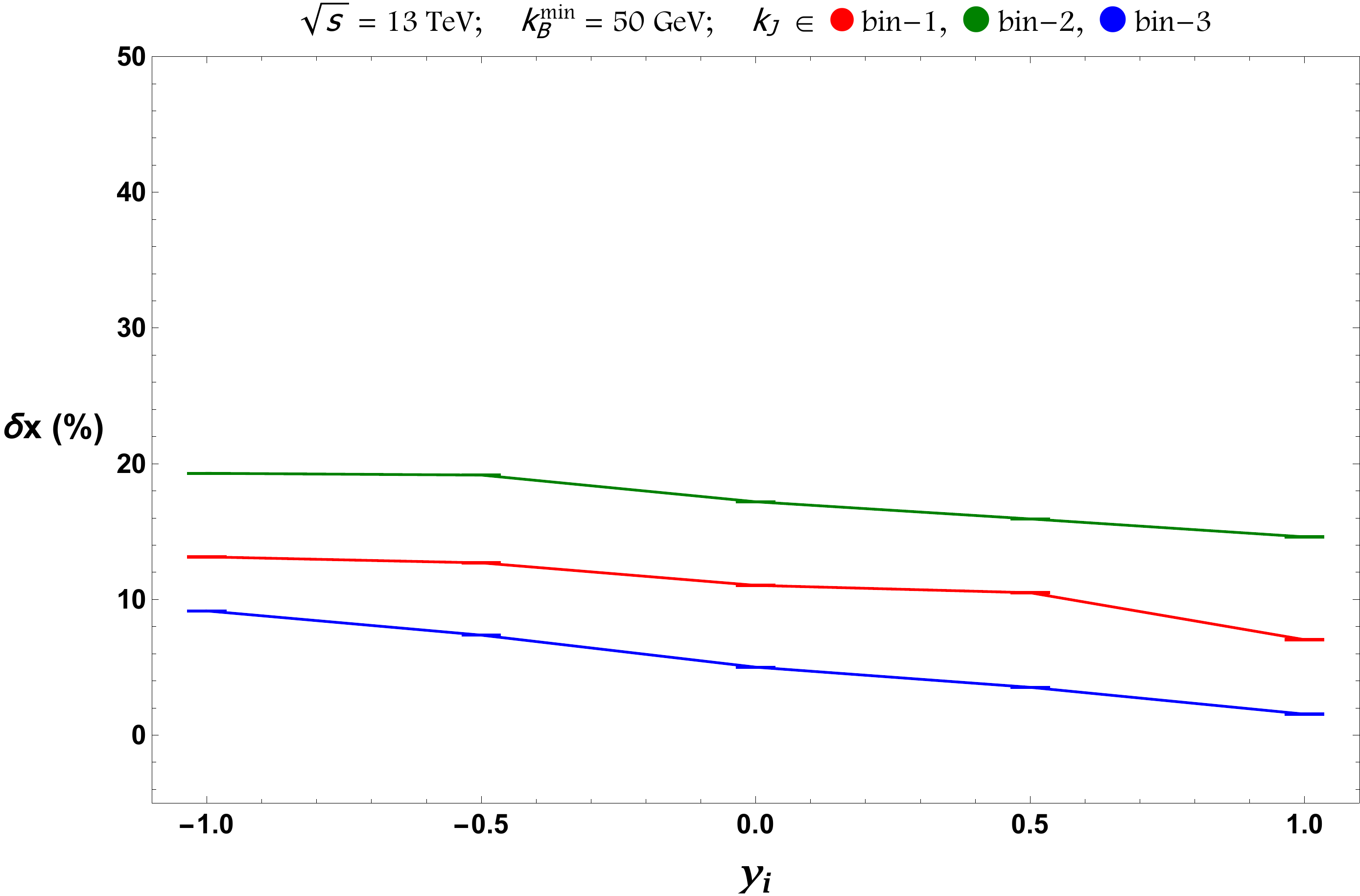}
   \vspace{1cm}

   \hspace{-16.25cm}   
   \includegraphics[scale=0.3]{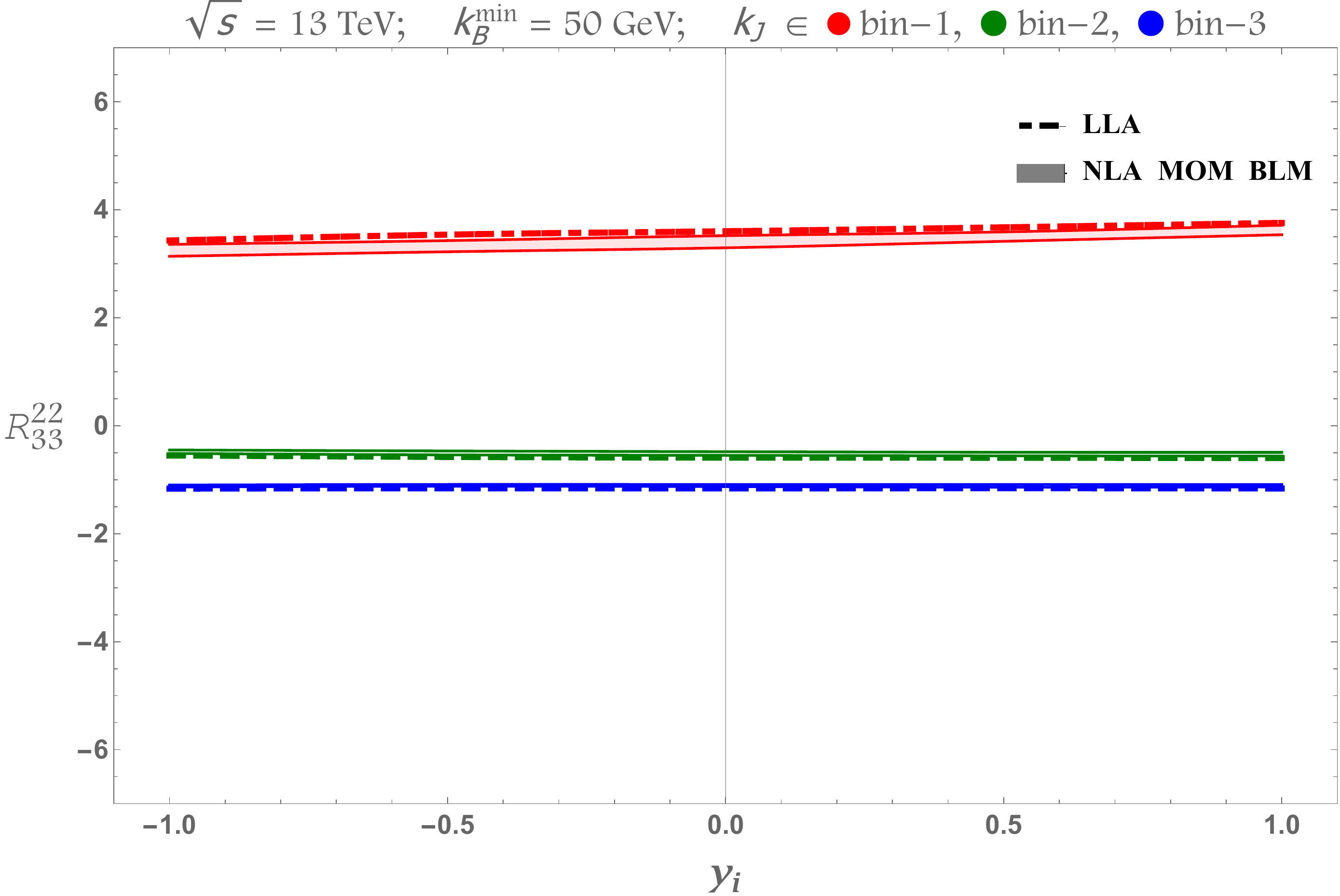}
   \includegraphics[scale=0.3]{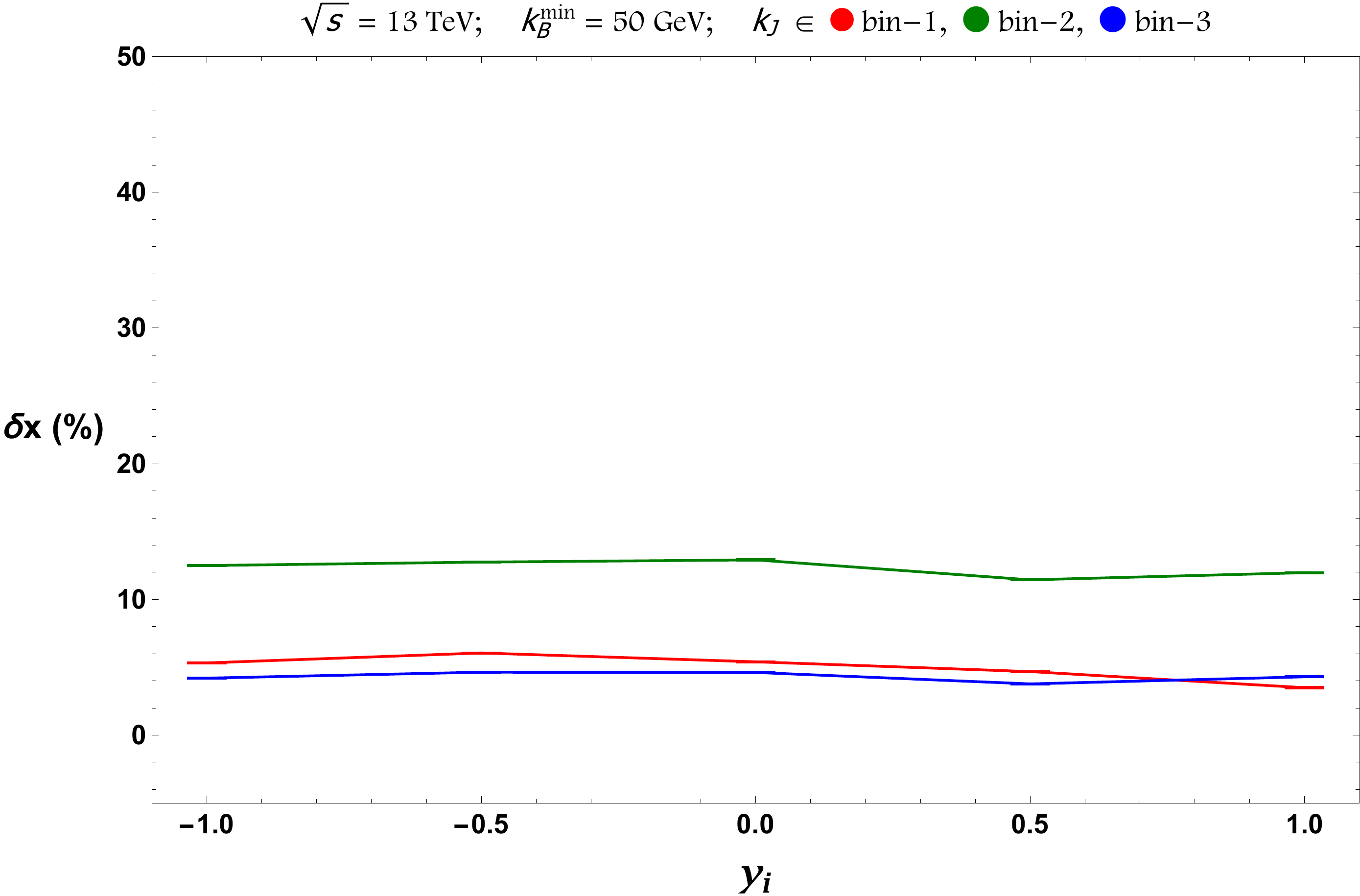}

\restoregeometry
\caption{\small $y_i$-dependence of the LLA and NLLA
$R^{12}_{22}$, $R^{12}_{33}$ and $R^{22}_{33}$ 
at $\sqrt s = 13$ TeV (left) and the relative NLLA to LLA corrections (right).} 
\label{fig:13-third}
\end{figure}

Again, keeping our notation with regard to the ratios uniform, we continue 
denoting our observables by $R_{PQ}^{MN}$
but now the ratios are  functions of $y_i$ instead of $Y$:
\begin{eqnarray}
\label{RPQMN}
R_{PQ}^{MN}(y_i) \, = \, \frac{C_{MN}^{\text{integ}}(y_i)}{C_{PQ}^{\text{integ}}(y_i)}\,.
\label{RmnqpNew2}
\end{eqnarray}
We present our results in Figs.~\ref{fig:7-third} and~\ref{fig:13-third}.
We see that indeed, the $y_i$-dependence of the three ratios is very weak.
Moreover, the similarity between the $\sqrt s = 7$ TeV and $\sqrt s = 13$ TeV
plots is more striking that in the previous sections.
The relative NLLA to LLA corrections seem to be slightly larger here than in the previous sections.
We would like to stress once more that the results in this section are readily
comparable to the experimental data once the same cuts  are applied in the 
experimental analysis.

\section{Summary \& Outlook}

We have presented a first complete phenomenological study beyond the LLA of inclusive
three-jet production at the LHC within the BFKL framework, focussing on  azimuthal-angle dependent observables. 
We considered two colliding
energies, $\sqrt s = 7, 13$ TeV 
and an asymmetric kinematic cut with respect to the transverse momentum
of the forward ($k_A$) and backward ($k_B$) jets. 
In addition, we have chosen to consider an
extra condition regarding the value of the transverse momentum $k_J$ of the central jet,
dividing the allowed region for $k_J$ into three sub-regions: $k_J$ smaller than
$k_{A,B}$, $k_J$ similar to $k_{A,B}$ and $k_J$ larger than $k_{A,B}$.

For a proper study at full NLLA, one needs to consider the NLO jet vertices
and the NLLA gluon Green functions. We have argued that we expect the
latter to be of higher relevance and we proceed to calculate them
using the BLM prescription which has been successful in previous phenomenological analyses.
We have shown how our observables 
$R^{12}_{22}$, $R^{12}_{33}$ and $R^{22}_{33}$
change when we vary the rapidity
difference Y between $k_A$ and $k_B$ from 5.5 to 9 units for a fixed $y_J$ and
from 6.5 to 9 units for $-0.5 < y_J < 0.5$. We have presented
both the LLA and NLLA results along with plots that show the relative size of the 
NLLA corrections compared to the LLA ones.
We have also presented an alternative kinematical setup where
we allow for $Y_A$ and $Y_B$
to take
values such that $3 < Y_A <  4.7$ and
$-4.7 < Y_B < -3$, while
the rapidity of the central jet takes values in five distinct rapidity bins of unit width, that is,
$y_i-0.5 < y_J<y_i+0.5$, with $y_i = \{-1, -0.5, 0, 0.5, 1\}$. 
In this alternative setup,
we presented
our results for $R_{12}^{22}$, $R_{12}^{33}$ and $R_{22}^{33}$ 
as functions of $y_i$.

The general conclusion is that the NLLA corrections are moderate and our proposed
observables exhibit a good perturbative stability. Furthermore, we see that for a wide
range of rapidities, the changes we notice when going from 7 TeV to 13 TeV are
small which makes us confident that
these generalised ratios pinpoint the crucial characteristics
of the BFKL dynamics  regarding
the azimuthal behavior of the hard jets in inclusive three-jet production.
It will be very interesting to compare with possible predictions for these observables from 
fixed order analyses as well as from the BFKL inspired 
Monte Carlo \cod {BFKLex}~\cite{Chachamis:2011rw,Chachamis:2011nz,Chachamis:2012fk,
Chachamis:2012qw,Caporale:2013bva,Chachamis:2015zzp,Chachamis:2015ico,Chachamis:2016ejm}. 
Predictions from general-purpose Monte Carlos tools should also be welcome.
It would be extremely interesting to pursue an
experimental analysis for these observables using  LHC data.

\begin{flushleft}
{\bf \large Acknowledgements}
\end{flushleft}
GC acknowledges support from the MICINN, Spain, 
under contract FPA2013-44773-P. 
ASV acknowledges support from Spanish Government 
(MICINN (FPA2010-17747,FPA2012-32828)) and, together with FC and FGC, 
to the Spanish MINECO Centro de Excelencia Severo Ochoa Programme 
(SEV-2012-0249). DGG is supported with a fellowship of the international programme "La Caixa-Severo Ochoa".
FGC thanks the Instituto de F{\'\i}sica Te{\'o}rica 
(IFT UAM-CSIC) in Madrid for warm hospitality.

\appendix
\section{$y_J$ independent integrated distributions}
We show now how Eq.~20 is fulfilled in our normalisations.
 We introduce the notation $t = \ln{k^2}$ to write the gluon Green function in the form
\begin{eqnarray}
\varphi \left(t_A,t_B,\theta_A,\theta_B,Y\right)  &=& \frac{e^{-\frac{t_A+t_B}{2}} }{\pi^2} 
\sum_{n=-\infty}^\infty e^{i n \left(\theta_A - \theta_B\right)} \nonumber\\
& & \int_0^\infty d \nu   
\cos{\left(\nu \left(t_A-t_B\right)\right)} \, e^{\bar{\alpha}_s  \chi_{|n|} \left(\nu\right) Y}.
\end{eqnarray}
Making use of $d k = \frac{1}{2} e^{\frac{t}{2}} dt$ and $k \, dk \, d \theta = \frac{e^t}{2} d \theta$ we then want to 
show that
\begin{eqnarray}
\varphi \left(t_A,t_B,\theta_A,\theta_B,Y\right) & = &
  \int_0^{2 \pi} d \theta 
\int_{-\infty}^\infty dt \, \frac{ e^t}{2} 
\varphi \left(t_A,t,\theta_A,\theta,y\right)\varphi \left(t,t_B,\theta,\theta_B,Y-y\right) \nonumber\\
&&\hspace{-3.5cm} = \int_0^{2 \pi} d \theta 
\int_{-\infty}^\infty dt \, \frac{ e^t}{2} \frac{e^{-\frac{t_A+t}{2}} }{\pi^2} 
\sum_{m=-\infty}^\infty e^{i m \left(\theta_A - \theta\right)}
\int_0^\infty d \nu   
\cos{\left(\nu \left(t_A-t\right)\right)} \, e^{\bar{\alpha}_s  \chi_{|m|} \left(\nu\right) y} \nonumber\\
&&\hspace{-3.cm} 
 \frac{e^{-\frac{t+t_B}{2}} }{\pi^2} 
\sum_{n=-\infty}^\infty e^{i n \left(\theta - \theta_B\right)}
\int_0^\infty d \mu   
\cos{\left(\mu \left(t-t_B\right)\right)} \, e^{\bar{\alpha}_s  \chi_{|n|} \left(\mu\right) (Y-y)}.
\end{eqnarray}
The integration over $\theta$ generates a $\delta_m^n$ contribution:
\begin{eqnarray}
\varphi \left(t_A,t_B,\theta_A,\theta_B,Y\right) &=&\nonumber\\
&&\hspace{-3.5cm}   \frac{e^{-\frac{t_A+t_B}{2}} }{\pi^3}  \sum_{n=-\infty}^\infty 
 e^{i n  \left(\theta_A - \theta_B\right)}
\int_0^\infty d \nu   
\, e^{\bar{\alpha}_s  \chi_{|n|} \left(\nu\right) y} \nonumber\\
&&\hspace{-3.5cm} 
\int_0^\infty d \mu   
 \, e^{\bar{\alpha}_s  \chi_{|n|} \left(\mu\right) (Y-y)}
\int_{-\infty}^\infty dt  \cos{\left(\nu \left(t_A-t\right)\right)} \cos{\left(\mu \left(t-t_B\right)\right)}.
\end{eqnarray}

It can be shown that
\begin{eqnarray}
\int_{-\infty}^\infty dt  \cos{\left(\nu \left(t_A-t\right)\right)} \cos{\left(\mu \left(t-t_B\right)\right)} &=& \nonumber\\
&&\hspace{-6cm} \pi 
\Bigg( \cos{ (\nu t_A - \mu t_B)}  \delta (\nu-\mu)
+ \cos{ ( \nu t_A+ \mu t_B)} \delta (\nu+\mu)
 \Bigg), 
\end{eqnarray}
which can be used to write Eq.~25 as
\begin{eqnarray}
\varphi \left(t_A,t_B,\theta_A,\theta_B,Y\right) &=&\nonumber\\
&&\hspace{-3.5cm}   \frac{e^{-\frac{t_A+t_B}{2}} }{2 \pi^2}  \sum_{n=-\infty}^\infty 
 e^{i n  \left(\theta_A - \theta_B\right)}
\int_{-\infty}^\infty d \nu   
\int_0^\infty d \mu   
 \, e^{\bar{\alpha}_s  \chi_{|n|} \left(\mu\right) (Y-y)} e^{\bar{\alpha}_s  \chi_{|n|} \left(\nu\right) y}  \nonumber\\
&&\hspace{-2.cm} 
\Bigg( \cos{ (\nu t_A - \mu t_B)}  \delta (\nu-\mu)
+ \cos{ ( \nu t_A+ \mu t_B)} \delta (\nu+\mu)
 \Bigg),
\end{eqnarray}
and, finally,
\begin{eqnarray}
\varphi \left(t_A,t_B,\theta_A,\theta_B,Y\right) &=&\nonumber\\
&&\hspace{-3.5cm}  \frac{e^{-\frac{t_A+t_B}{2}} }{\pi^2}  \sum_{n=-\infty}^\infty 
 e^{i n  \left(\theta_A - \theta_B\right)}   
\int_0^\infty d \mu   
 \, e^{\bar{\alpha}_s  \chi_{|n|} \left(\mu\right) Y}\cos{ (\mu (t_A -t_B))},
\end{eqnarray}
which is the same as our initial representation for $\varphi$ in Eq.~23.
The relation in Eq.~24 is remarkable because it holds for 
any rapidity $y$.

\end{document}